\documentclass[%
 reprint,
 amsmath,amssymb,
 aps,
]{revtex4-2}

\usepackage{graphicx}
\usepackage{dcolumn}
\usepackage{bm}

\begin{document}

\title{Quantum multiparameter estimation with multi-mode photon catalysis entangled squeezed state}
\author{Huan Zhang$^{1}$}
\author{Wei Ye$^{2}$}
\author{Shoukang Chang$^{3}$}
\author{Ying Xia$^{1}$}
\author{Liyun Hu$^{4}$}
\thanks{hlyun@jxnu.edu.cn}
\author{Zeyang Liao$^{1}$}
\thanks{liaozy7@mail.sysu.edu.cn}
\affiliation{$^{{\small 1}}$\textit{School of Physics, Sun Yat-sen University, Guangzhou 510275, China}\\
$^{{\small 2}}$\textit{School of Information Engineering, Nanchang Hangkong University, Nanchang 330063, China}\\
$^{{\small 3}}$\textit{MOE Key Laboratory for Nonequilibrium Synthesis and Modulation of Condensed Matter, Shaanxi Province Key Laboratory of Quantum Information and Quantum Optoelectronic Devices, School of Physics, Xi'an Jiaotong University, 710049, People's Republic of China}\\
$^{{\small 4}}$\textit{Center for Quantum Science and Technology, Jiangxi Normal University, Nanchang 330022, China}}

\begin{abstract}
We propose a method to generate the multi-mode entangled catalysis squeezed vacuum states (MECSVS) by embedding the cross-Kerr nonlinear medium into the Mach-Zehnder interferometer. This method realizes the exchange of quantum states between different modes based on Fredkin gate. In addition, we study the MECSVS as the probe state of multi-arm optical interferometer to realize multi-phase simultaneous estimation. The results show that the quantum Cramer-Rao bound (QCRB) of phase estimation can be improved by increasing the number of catalytic photons or decreasing the transmissivity of the optical beam splitter using for photon catalysis.  In addition, we also show that even if there is photon loss, the QCRB of our photon catalysis scheme is lower than that of the ideal entangled squeezed vacuum states (ESVS), which shows that by performing the photon catalytic operation is  more robust against photon loss than that without the catalytic operation. The results here can find important applications in quantum metrology for multiparatmeter estimation. 

\textbf{PACS: }03.67.-a, 05.30.-d, 42.50,Dv, 03.65.Wj
\end{abstract}

\maketitle

\section{Introduction}
Quantum metrology is one of the most important research fields in quantum information science which can provide significant quantum advantages over its classical counterpart. A fundamental task in quantum metrology is to improve the estimation precision of parameters to be measured through quantum resources allowing by the basic principles of quantum mechanics. Generally speaking, the typical quantum metrology includes three steps: the preparation of probe states, the evolution of probe states, and the readout of the evolved states. In quantum metrology, the quantum Cramer-Rao bound (QCRB) is usually used to  quantify the estimation precision offered by quantum metrology, which gives the lower limit of the estimation precision that can be achieved using any possible detection methods \cite{1,2,3,4,5}. For this purpose, prior works are focused on the estimation of a single parameter with superior quantum resources \cite{6,7,8,9,10,11,12,13,14,15}. For instance, by adopting nonclassical or entangled states, such as single (two)-mode squeezed vacuum state (S(T)SVS) \cite{11,12,12a} and NOON state \cite{13}, the estimation precision can overcome the so-called standard quantum limit (SQL) scaling as $1/\sqrt{N}$ with $N$ being the mean photon number of the probe state, and in certain cases, the precision can even approach to the renowned Heisenberg limit (HL) with a scaling $1/N$.  Although theoretically we can achieve arbitrary large squeezing parameter, in practice increasing the squeezing parameter is not an easy task \cite{13a}. If we can perform certain non-Gaussian operations such as photon addition, subtraction and catalysis on these experimental achievable squeezing states, we may further enhance the precision of the metrology and achieve the same precision as those by inputting a higher squeezing state which is currently  hard to be generated in experiment \cite{14,15,16,17,18,42a}.

In the realistic scenarios, such as biological system measurement \cite{19,20,21}, the optical imaging and the sensor network \cite{22,23,24,25,26}, multiparameter quantum metrology is indispensable and thus has received a lot of increasing interest in recent years \cite{27,28,29,31,31a,31b,32,33,34,35}, as the number of parameters affecting a physical process is usually more than one. For instance, Humphreys \textit{et al}. treated the phase imaging problem regarded as a multiparameter estimation process, and showed the advantages of the multiparameter simultaneous estimation using the multi-mode NOON state, when comparing with the independent estimation scheme \cite{29}. The advantages remain even if  there are photon losses, as studied by Yue  \textit{et al}. \cite{36}. Apart from the photon losses, Ho \textit{et al}. estimated three components of an external magnetic field using the entangled Greenberger-Horne-Zeilinger state including the dephasing noise and showed that its sensitivity can beat the SQL \cite{37}. Besides, Yao \textit{et al}. investigated the multiple phase estimation problem for a natural parametrization of arbitrary pure states under the white noise \cite{38}. In order to further develop the quantum enhanced multiparameter simultaneous estimation, Hong \textit{et al}. proposed a method to generate the multi-mode NOON state, and they experimentally demonstrated that the QCRB can be saturated using the multi-mode NOON state \cite{39}. In addition to the NOON state, entangled coherent state is also widely used in the field of quantum metrology. To compare the performance of the entangled coherent state can be better with that of the multi-mode NOON state \cite{3,10,27,40,41}, Liu \textit{et al}. proposed a theoretical scheme of quantum enhanced multiparameter metrology with generalized entangled coherent state and showed that the entangled coherent state can indeed give better precision than that of the multi-mode NOON state \cite{27}. After that, Zhang \textit{et al.} investigated the quantum multiparameter estimation with generalized balanced multimode NOON-like states, including the entangled squeezed vacuum state (ESVS), the entangled squeezed coherent state, the entangled coherent state, and the NOON state \cite{42,49}. Comparing with other multimode NOON-like states, they found that the ESVS has the lowest QCRB if the mean photon number is the same.

In this paper, we propose a scheme to generate the multi-mode entangled catalysis squeezed vacuum states (MECSVS). Due to the fact that the multiphoton catalysis operation \cite{43,44} can improve the fidelity in quantum teleportation \cite{45,46}, extend the transmission distance in continuous variable quantum key distribution \cite{47,48}, and enhance the sensitivity of phase estimation for a single-phase estimation \cite{18} and undo the noise effect of the channel \cite{48a}, we also propose ascheme to improve the multiparameter estimation precision by using the MECSVS. Our results clearly show that the multiphoton catalytic operation can further improve the precision of phase estimation compared with the result with ordinary ESVS as the probe state. Moreover, the usage of multi-photon catalysis in the multiparameter quantum metrology is also more robust against the photon losses which can find important applications in the practical scenarios.


The paper is organized as follows: In Sec. II, we propose a scheme to generate the MECSVS. In Sec. III and IV, we evaluate the QCRB of multiparameter estimation with the symmetric MECSVS under ideal and photon-loss cases, respectively. In Sec. V, we study the QCRB when the anti-symmetric MECSVS is used. Finally, we summarize the results. 

\section{The generation of the MECSVS}

\begin{figure*}[tbp]
\label{Fig1} \centering \includegraphics[width=16cm]{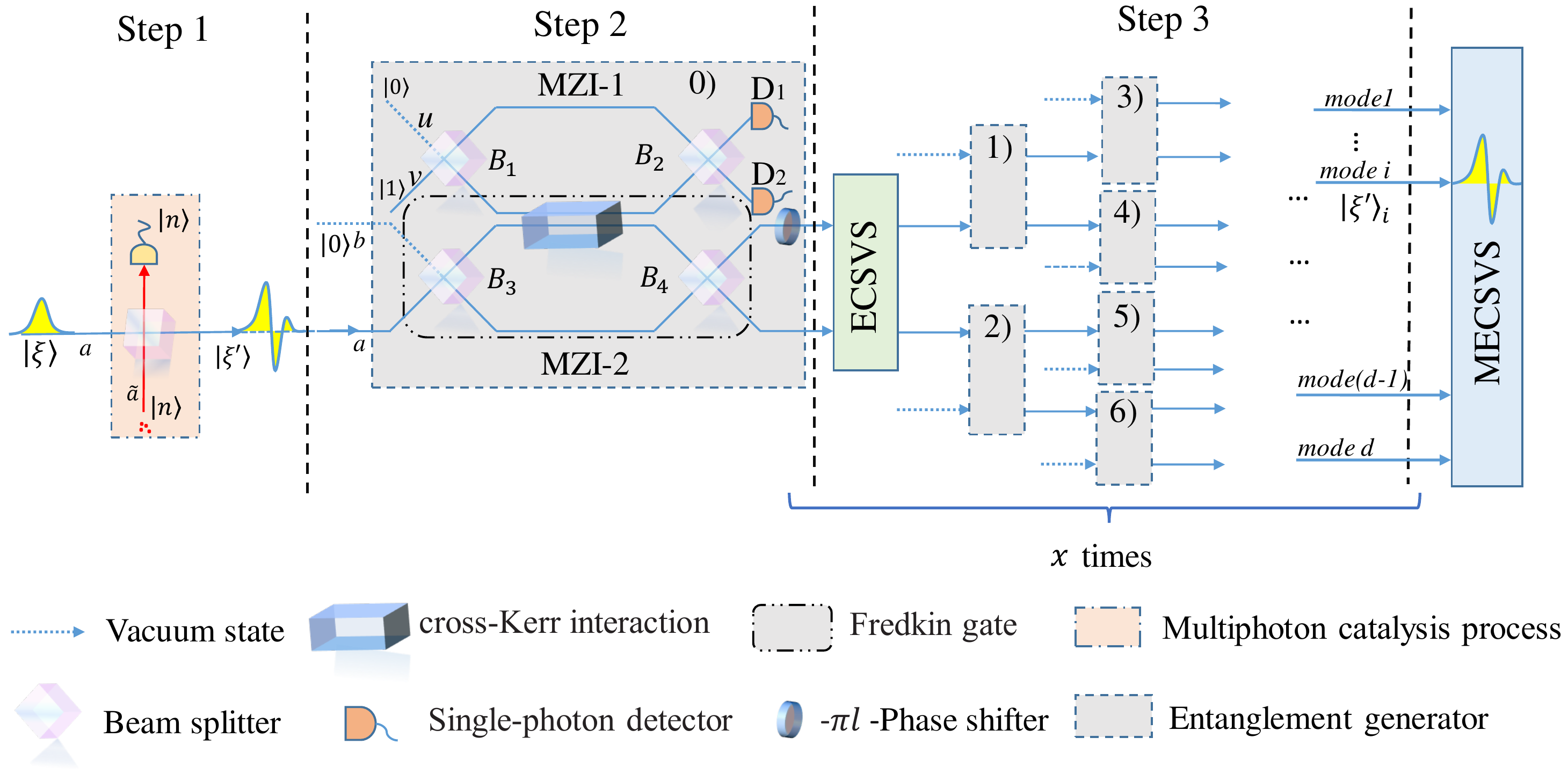}
\caption{{}Generation of the $d$-mode MECSVS using quantum-optical Fredkin gate. The orange box is the multiphoton catalysis process, in which an $n$-photon Fock state is inputted into auxiliary mode of beam splitter, conditional measuring $n$ photons at one output. The dashed box is the entanglement generator, which creates two-mode entangled catalysis squeezed vacuum state by using the single-mode multiphoton catalysis\ squeezed vacuum state and vacuum state. Applying the entanglement generator $x$ times with each input of the following generator aligning with one of the output modes of the previous generator creates a $d$-mode MECSVS. }
\end{figure*}

In this section, we propose a scheme to generate the MECSVS which is the probe quantum state used for the quantum multiparameter estimation. The schematic diagram is shown in Fig. 1 in which three steps are included. In the first step, we perform the $n$-photon catalysis (orange box) on the input SSVS $|\xi\rangle$. In the $n$-photon catalysis, an $n$-photon Fock state $\left\vert n\right\rangle $ in an ancillary mode $\overset{\thicksim }{a}$ is injected into one port of a beam splitter with a transmissivity $T,$ and then is detected at the corresponding output of mode $\overset{\thicksim }{a}$. In the meantime, the input squeezed state $|\xi\rangle$ is injected into the other port $a$ of the beam splitter. If $n$ photons are detected at the output port $\overset{\thicksim }{a}$, the so called $n-$photon catalysis is performed and the output state of port $a$ is given by $|\xi'\rangle$ which is the single-mode $n-$photon catalyzed squeezed state.  This multi-photon catalyzed process can be regarded as an effective operator \cite{45}, i.e.,
\begin{eqnarray}
\widehat{O} &\equiv &\left\langle n\right\vert \widehat{B}\left( T\right)
\left\vert n\right\rangle  \notag \\
&=&\left.\frac{T^{n/2}}{n!}\frac{\partial ^{n}}{\partial \tau ^{n}}\Big \{\frac{\exp
\left [ \mu(\tau) a^{\dagger }a\right ] }{1-\tau }\Big \}\right\vert _{\tau =0},  \label{1}
\end{eqnarray}%
where $\widehat{B}\left( T\right) =e^{(\widetilde{a}^{\dagger}a-\widetilde{a}a^{\dagger })\arccos \sqrt{T}}$ and the parameter
\begin{equation}
\mu (\tau) =\ln \frac{\sqrt{T}-\tau /\sqrt{T}}{1-\tau }.  \label{2}
\end{equation}%
Considering that the input state of $a$-mode is a Gaussian SSVS $\left\vert
\xi \right\rangle =\sqrt{sech r}\sum_{l=0}^{\infty }\left( -\frac{1}{2}\tanh r\right)^{l}\sqrt{(2l)!}/l!\left\vert 2l\right\rangle $ as the input state, the corresponding single-mode multiphoton catalysis\ squeezed vacuum state is thus given by
\begin{eqnarray}
\left\vert \xi ^{\prime }\right\rangle &=&%
\frac{1}{\sqrt{\mathbb{N}}}
\widehat{O}\left\vert \xi \right\rangle _{a}  \notag \\
&=&%
\left.\frac{1}{\sqrt{\mathbb{N}}}
\frac{T^{n/2}}{n!}\frac{\partial ^{n}}{\partial \tau ^{n}} \Big [\beta(\tau)
\sum_{l=0}^{\infty }c_{l}(\tau)\left\vert 2l\right\rangle _{a} \Big ]\right\vert _{\tau =0},  \label{3}
\end{eqnarray}%
with $\beta(\tau) =1/\left( 1-\tau \right) \sqrt{\cosh r,}$ $c_{l}(\tau)=\left[ -\frac{1}{%
2}e^{2\mu(\tau) }\tanh r\right] ^{l}\sqrt{\left( 2l\right) !}/l!$ and $%
\mathbb{N}
$ is the normalization factor given by
\begin{equation}
\mathbb{N}
=\left.\frac{T^{n}}{\left( n!\right) ^{2}}\frac{\partial ^{2n}}{\partial \tau
^{n}\partial \tau ^{* n}}\aleph \left( \tau ,\tau ^{* }\right)
\big[ 1-\Im \left( \tau ,\tau ^{* }\right) \big] ^{-\frac{1}{2}%
}\right\vert _{\tau =\tau ^{* }=0},  \label{4}
\end{equation}%
with
\begin{eqnarray}
\aleph \left( \tau ,\tau ^{* }\right)  &=&\frac{1}{\left( 1-\tau
\right) \left( 1-\tau ^{* }\right) \cosh r},  \notag \\
\Im \left( \tau ,\tau ^{* }\right)  &=&e^{2[ \mu (\tau^*)+\mu (\tau)]}\tanh ^{2}r,  \label{5}
\end{eqnarray}
where $\tau ^{* }$ is the complex conjugate of $\tau $. In particular, from Eq. (\ref{3}), when $T=1$, the single-mode multiphoton catalysis squeezed vacuum state is reduced to the SSVS,
as expected. Although the photon catalysis operation is probabilistic  \cite{49a}, it can be heralded and we choose to perform the quantum metrology only when the MECSVS is successfully generated.

After generating the single-mode multi-photon catalysis squeezed vacuum state as shown in Eq.(\ref{3}), we next produce an entangled squeezed state using step 2 as shown in Fig. 1. For this purpose, we propose to employ that quantum-optical Fredkin gate which consists of two Mach-Zehnder interferometers mediated by a cross-Kerr medium.  To be more specific, the single-mode multiphoton
catalysis squeezed vacuum states and a vacuum state $\left\vert0\right\rangle $ are respectively injected into a cross-Kerr medium based Mach-Zehnder interferometer (MZI-2) from modes $a$ and $b$. Simultaneously, a vacuum state $\left\vert 0\right\rangle _{u}$ and a single photon state $\left\vert 1\right\rangle _{v}$ are used as inputs of the MZI-1. We assume that the four beam splitters $\widehat{B}_{j}$ $(j=1,2,3,4)$ are chosen as $50:50$, i.e., $\widehat{B}_{1}=e^{i\pi \left( u^{\dagger }v+uv^{\dagger }\right) /4}$ and $\widehat{B}_{3}=e^{i\pi \left( a^{\dagger }b+ab^{\dagger }\right) /4}$ with $\widehat{B}_{4}=\widehat{B}_{3}^{\dagger }$ and $\widehat{B}_{2}=\widehat{B}_{1}^{\dagger }$ . After passing through the beams splitters $B_1$ and $B_3$, the photons in mode $a$ and mode $u$ pass through the corss-Kerr medium at the same time and the effective operator is given by \cite{49,50}
\begin{equation}
\widehat{U}_{k}=\exp \left( i\chi t a^{\dagger }au^{\dagger }u\right) ,
\label{6}
\end{equation}%
where $\chi$ is the nonlinear Kerr coupling coefficient and $t$ is the interaction time. For our purpose here, we choose  $\chi t =\pi $. Unfortunately, the cross-Kerr nonlinearity in the natural medium is usually very small and $\chi t$ is usually much less than $\pi $. However, enhancement of the cross-Kerr nonlinearity is not impossible. In the past few decades, a number of methods have been proposed to achieve giant Kerr nonlinearity. For example, giant Kerr-nonlinearity such that $\chi > 2\pi\times10^{10} Hz$ can be achieved in an atomic ensemble by the electromagnetic induced transparency (EIT) and the slow light effect  \cite{51,52,53,54,55}. Besides, in \cite{56}, the author showed that by constructing a one-dimensional nonlinear photonic crystal from alternating layers of Kerr medium and linear dielectric medium, the phase of the wave function of the incident photons can be rotated by $\pi$ phase. In addition, by measurement-induced quantum operations, Costanzo et al. demonstrated an experimental implementation of a strong Kerr nonlinearity where a $\pi$ phase shift is realized \cite{57}. Therefore, $\pi$ phase shift by the Kerr-type interaction is possible.

It is clearly seen that if the $u$-mode is vacuum, no phase shift for the $b$ mode. However, if the $u$-mode has 1 photon, there is a phase shift for the $b$ mode. Due to the phase shift, the $b$-mode may enter the upper path or lower path after passing through $B_4$. This is the basic principle for generating the entangled state in this setup. The combination of the operation $\widehat{U}_{F}=\widehat{B}_{4}\widehat{U}_{k}\widehat{B}_{3}$ is actually the quantum-optical Fredkin gate which is given by
\begin{equation}
\widehat{U}_{F}=e^{i\frac{\pi }{2}u^{\dagger }u\left( a^{\dagger
}a+b^{\dagger }b\right) }e^{\frac{\pi }{2}u^{\dagger }u\left( ab^{\dagger
}-a^{\dagger }b\right) }.  \label{8}
\end{equation}%
This quantum gate can effectively entangle the two photon modes. 

To be more specific, when $\left\vert \xi ^{\prime }\right\rangle $
and $\left\vert 0\right\rangle $ are injected into the MZI-2 ($\left\vert
0\right\rangle _{u}$ and $\left\vert 1\right\rangle _{v}$ are injected into
the MZI-1), the unnormalized output state can be expressed as
\begin{eqnarray}
&&\widehat{B}_{2}\widehat{B}_{4}\widehat{U}_{k}\widehat{B}_{3}\widehat{B}_{1}\left\vert\xi ^{\prime } \right\rangle
_{a}\left\vert 0\right\rangle _{b}\left\vert 0\right\rangle
_{u}\left\vert 1\right\rangle _{v}  \notag \\
&\text{=}&\frac{1}{2}[\left( \left\vert 0\right\rangle _{a}\left\vert
e^{i\pi l}\xi ^{\prime }\right\rangle _{b}+\left\vert \xi ^{\prime
}\right\rangle _{a}\left\vert 0\right\rangle _{b}\right) \left\vert
1\right\rangle _{u}\left\vert 0\right\rangle _{v}  \notag \\
&&+i\left( -\left\vert 0\right\rangle _{a}\left\vert e^{i\pi l}\xi ^{\prime
}\right\rangle _{b}+\left\vert \xi ^{\prime }\right\rangle _{a}\left\vert
0\right\rangle _{b}\right) \left\vert 0\right\rangle _{u}\left\vert
1\right\rangle _{v}].  \label{9}
\end{eqnarray}%
Here, the state $\left\vert e^{i\pi l}\xi ^{\prime }\right\rangle _{b}$ describes the state similar to Eq. (\ref{3}) but the coefficient $c_{l}(\tau)$ is multiplied by a phase factor $e^{i\pi l}$. To remove this additional phase shift, we place a quarter wave plate in the output route of $b$ mode. When $2l$ photons pass through this quarter wave plate, an additional $\pi l$ phase shift is accumulated which exactly cancels out the previous phase factor. Two single photon detectors $D_1$ and $D_2$ are placed in the output routes of $u$ and $v$ modes. According to Eq. (\ref{9}), the output state depends on the detection results of $u$ and $v$ modes. Finally, the output state is given by
\begin{equation}
\left\vert \psi \right\rangle _{ab} = \frac{1}{\sqrt{\widetilde{N}_2}} (\pm\left\vert0 \right\rangle _{a}\left\vert \xi ^{\prime
}\right\rangle _{b}+\left\vert \xi ^{\prime }\right\rangle
_{a}\left\vert0 \right\rangle _{b}),  \label{10}
\end{equation}%
where $\widetilde{N}_2$ is a normalization factor. The symmetric (antisymmetric) state is obtained when a single photon is detected in $D_1$ ($D_2$ ) and no photon is detected in $D_2$ ($D_1$ ). It is clearly seen that the two-mode ECSVS is generated. Both the symmetric and the antisymmetric  ECSVSs can be used for improving the precision of the metrology. In the following, we mainly consider the symmetric case and disscuss the antisymmetric case in Sec. V.

After preparing the two-mode ECSVS state, we use it as inputs of two MZIs and repeat the procedures as those in step 2. By repeating these procedures for a number of time, we can in principle generate the MECSVS. If all the detection results of the auxiliar qubits are one photon in the  mode $u$ and zero photon in the  mode $v$, the output state is then given by 
\begin{equation}
\left\vert \Psi \right\rangle =\frac{1}{\sqrt{\widetilde{%
\mathbb{N}
}}}\sum_{j=0}^{d}\left\vert 0\right\rangle _{0}\left\vert 0\right\rangle
_{1}\left\vert 0\right\rangle _{2}\left\vert 0\right\rangle
_{3}...\left\vert \xi ^{\prime }\right\rangle _{j}...\left\vert
0\right\rangle _{d},  \label{11}
\end{equation}%
where $\widetilde{%
\mathbb{N}
}=1/\sqrt{\left( d+1\right) \left( \left\langle \xi ^{\prime }\right\vert
\left. \xi ^{\prime }\right\rangle +d\left\vert \left\langle 0\right\vert
\left. \xi ^{\prime }\right\rangle \right\vert ^{2}\right) }$ is the normalized coefficient, which can be calculated as%
\begin{eqnarray}
\widetilde{%
\mathbb{N}
} &=&\frac{T^{n}}{\left( n!\right) ^{2}}\frac{\partial ^{2n}}{\partial \tau
^{n}\partial \tau ^{* n}}\left( d+1\right) \aleph \left( \tau ,\tau ^{* }\right)   \notag \\
&&\left.\times \left\{ \left[ 1-\Im \left( \tau ,\tau ^{* }\right) \right]
^{-1/2}+d\right\} \right\vert _{\tau =\tau ^{* }=0}.  \label{12}
\end{eqnarray}
The quantum state shown in Eq. (\ref{11}) is a symmetric MECSVS which is a highly entangled state. In particular,  when $T=1$, the MECSVS is reduced to the multi-mode ESVS, as expected. In the following, we shall use the MECSVS as the inputs of a multi-arm interferometer in order to effectively improve the precision of multiparameter estimations of multiple optical phases at the same time with and without photon losses. 

\section{The QCRB of multiparameter estimation without photon losses}

\begin{figure}[tbp]
\label{Fig2} \centering \includegraphics[width=7cm]{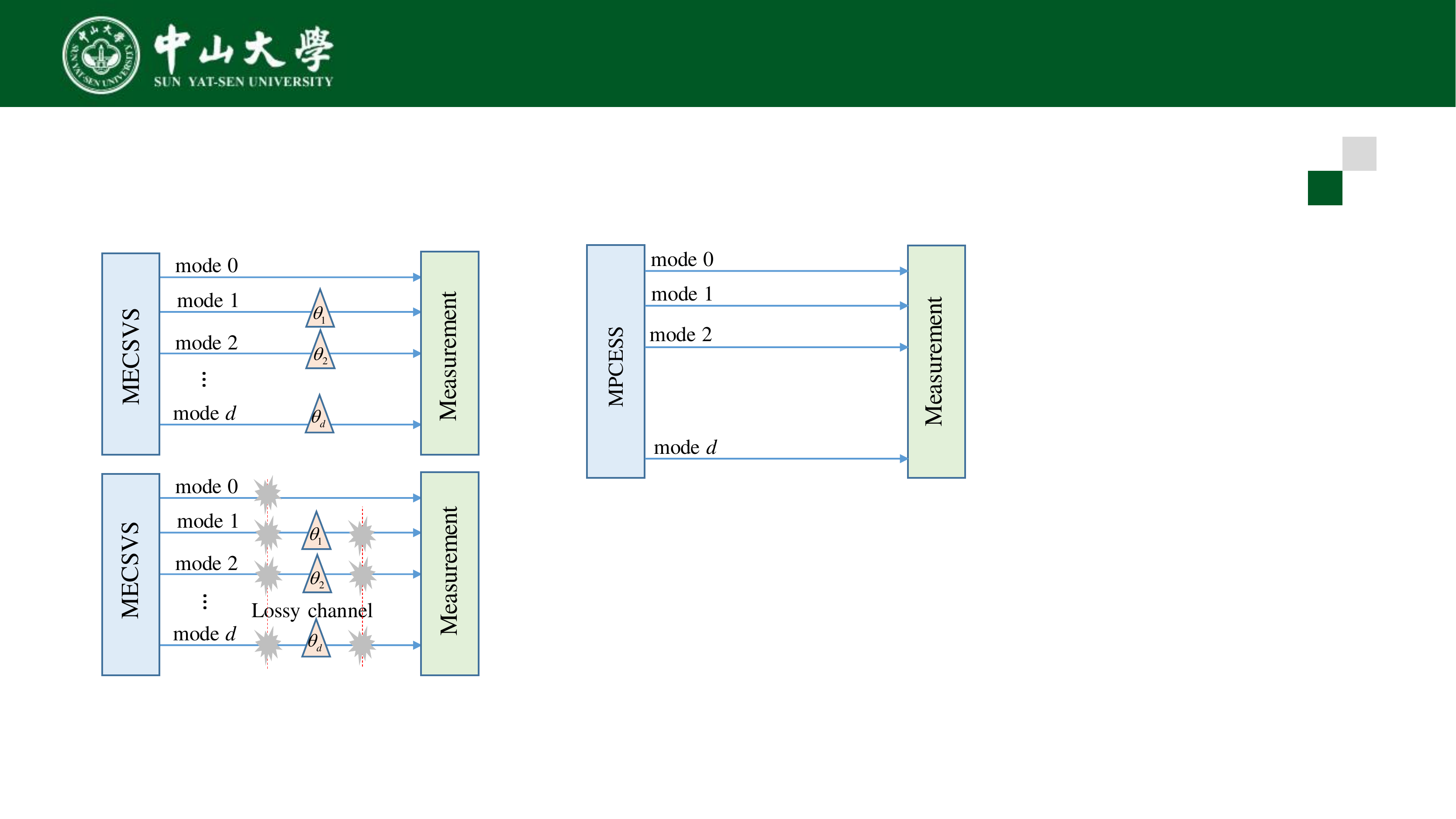}
\caption{{}(Color online) Schematic of multiparameter estimation with d phase shifts under an ideal case. }
\end{figure}

In this section, we investigate the QCRB of multiple independent phases estimation simultaneously by a multi-arm interferometer with inputting the MECSVS shown in Eq. (\ref{11}) under the ideal case. The schematic diagram of $d$ parameters estimation is shown in Fig. 2 where the mode $0$ is  the reference beam and the modes from $1$ to $d$ are the parametric beams.  The $d$ estimated parameters are assumed to be mutually independent linear phases and their transformations through the multi-arm interferometer can be represented by a unitary operator\cite{29,42}
\begin{equation}
\widehat{U}_{\theta }=\exp (i\sum_{j=1}^{d}\theta _{j}\cdot \widehat{N}_{j}),
\label{13}
\end{equation}%
where $\theta _{j}$ and $\widehat{N}_{j}=a_{j}^{\dagger }a_{j}$ represent the phase shift and the photon number operator for the $j$th mode, respectively. Since all the phases are independent to each other, the photon number operators $\widehat{N}_{j}$ for different $j$ are commutable. When the input MECSVS $\left\vert \Psi\right\rangle $ goes through the interferometer, the output state can be expressed as $\left\vert \Psi_{out}\right\rangle =\widehat{U}_{\theta }\left\vert \Psi \right\rangle $. According to the definition of QCRB, the estimation precision of $\widehat{\mathbf{\theta }}$ is inversely proportional to the quantum Fisher inforamtion (QFI) of the output state $\left\vert \Psi_{out}\right\rangle$, i.e.,
\begin{equation}
\left\vert \Delta \widehat{\theta }\right\vert ^{2}\geq \left\vert \Delta
\widehat{\theta }\right\vert _{QCRB}^{2}=\text{Tr}[\mathbf{F}_{\theta
}^{-1}],  \label{14}
\end{equation}%
where Tr$[\cdot ]$ represents the trace operation and $\mathbf{F}_{\theta
}^{-1}$ is the inverse of the $d\times d$ QFI matrix. 

Given an arbitrary pure state, it is possible to saturate the QCRB  if $\text{Im}
\left\langle \psi \right\vert L_{j}L_{k}\left\vert \psi \right\rangle =0$ is
satisfied for all $j,k$ and $\theta ,$ in which $L_{j(k)}$ is the
symmetric logarithmic derivative given by $L_{j(k)}$=$2\left( \left\vert
\partial _{j(k)}\Psi _{out}\right\rangle \left\langle \Psi _{out}\right\vert
+\left\vert \Psi _{out}\right\rangle \left\langle \partial _{j(k)}\Psi
_{out}\right\vert \right) $ with $\left\vert \partial _{j(k)}\Psi
_{out}\right\rangle $=$\partial _{j(k)}\left\vert \Psi _{out}\right\rangle
/\partial _{j(k)}\theta _{j(k)}$ \cite{27,58,59}. For our scheme, based on
Eqs. (\ref{11}) and (\ref{13}), the elements of the QFI matrix are given
by
\begin{equation}
f_{jk}=4(\left\langle \Psi \right\vert \widehat{N}_{j}\widehat{N}%
_{k}\left\vert \Psi \right\rangle -\left\langle \Psi \right\vert \widehat{N}%
_{j}\left\vert \Psi \right\rangle \left\langle \Psi \right\vert \widehat{N}%
_{k}\left\vert \Psi \right\rangle ).  \label{15}
\end{equation}%
As a result, the QFI matrix reads as \cite{60}%
\begin{equation}
\mathbf{F}_{\mathbf{\theta }}=4\frac{1}{\widetilde{%
\mathbb{N}
}}\left\langle \xi ^{\prime }\right\vert \widehat{N}^{2}\left\vert \xi
^{\prime }\right\rangle I-4\frac{1}{\widetilde{%
\mathbb{N}
}^{2}}\left\langle \xi ^{\prime }\right\vert \widehat{N}\left\vert \xi
^{\prime }\right\rangle ^{2}\bar{I},  \label{16}
\end{equation}%
where $I$ represents the identity matrix and $\bar{I}$ denotes the matrix
with the elements $\bar{I}_{jk}=1$ for all $j$ and $k$. From Eqs. (\ref{14}) and (\ref{16}),
we can finally obtain the expression of the QCRB for our scheme, i.e.,
\begin{equation}
\left\vert \Delta \widehat{\mathbf{\theta }}\right\vert _{QCRB}^{2} =
\frac{d}{4\left\langle \xi ^{\prime }\right\vert \widehat{N}%
^{2}\left\vert \xi ^{\prime }\right\rangle }\left( \widetilde{%
\mathbb{N}
}+\frac{1}{\Bbbk -d/\widetilde{%
\mathbb{N}
}}\right) ,  \label{17}
\end{equation}%
where $\Bbbk =\left\langle \xi ^{\prime }\right\vert \widehat{N}%
^{2}\left\vert \xi ^{\prime }\right\rangle /\left\langle \xi ^{\prime
}\right\vert \widehat{N}\left\vert \xi ^{\prime }\right\rangle ^{2}$ with
\begin{eqnarray}
\left\langle \xi ^{\prime }\right\vert \widehat{N}^{2}\left\vert \xi
^{\prime }\right\rangle  &=&\frac{T^{n}}{\left( n!\right) ^{2}}\frac{%
\partial ^{2n}}{\partial \tau ^{n}\partial \tau ^{* n}}\aleph \left(
\tau ,\tau ^{* }\right)   \notag \\
&&\left.\times \frac{\Im \left( \tau ,\tau ^{* }\right) \left[ \Im \left(
\tau ,\tau ^{* }\right) +2\right] }{\left[ 1-\Im \left( \tau ,\tau ^{* }\right) \right] ^{5/2}}\right\vert _{\tau =\tau ^{* }=0},  \notag \\
\left\langle \xi ^{\prime }\right\vert \widehat{N}\left\vert \xi ^{\prime
}\right\rangle  &=&\frac{T^{n}}{\left( n!\right) ^{2}}\frac{\partial ^{2n}}{%
\partial \tau ^{n}\partial \tau ^{* n}}\aleph \left( \tau ,\tau ^{* }\right)   \notag \\
&&\left.\times \frac{\Im \left( \tau ,\tau ^{* }\right) }{\left[ 1-\Im \left(
\tau ,\tau ^{* }\right) \right] ^{3/2}}\right\vert _{\tau =\tau ^{* }=0}.  \label{18}
\end{eqnarray}%
Note that $\left\vert \Delta \widehat{\mathbf{\theta }}\right\vert
_{QCRB}^{2}$ is positive definite according to Eq. (\ref{17}). Especially,
when $T=1$, one can obtain $\left\langle \xi ^{\prime }\right\vert \widehat{N%
}^{2}\left\vert \xi ^{\prime }\right\rangle =\sinh ^{4}r+2\sinh ^{2}r\cosh
^{2}r$ and $\left\langle \xi ^{\prime }\right\vert \widehat{N}\left\vert \xi
^{\prime }\right\rangle =\sinh ^{2}r$, which is consistent with the result of the $d+1$
modes ESVS case, as expected \cite{42}. When $T\neq 1$, the QCRB can be significantly different from that of the ESVS case.

\begin{figure*}[tbp]
\label{Fig3} \centering \includegraphics[width=0.68\columnwidth]{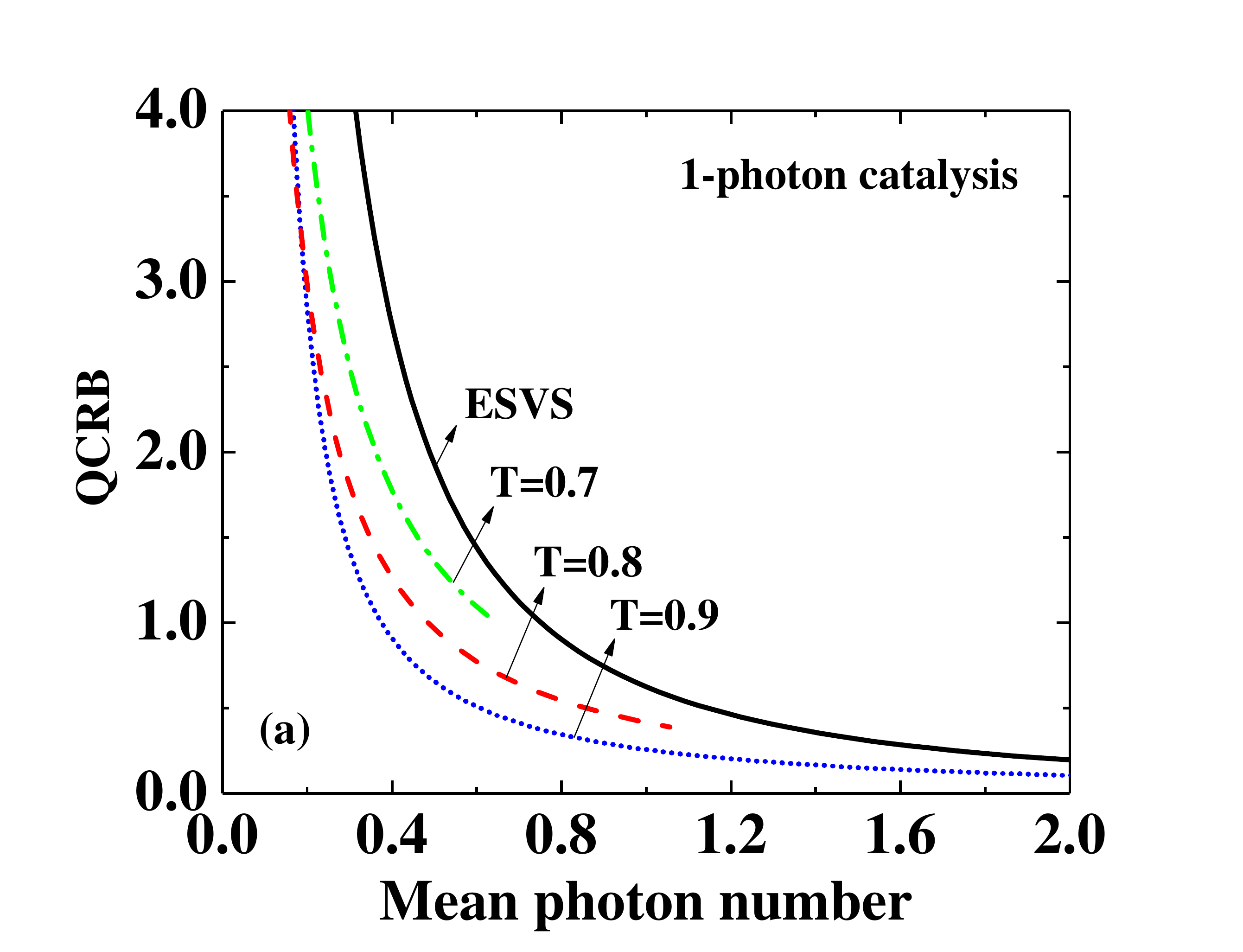} %
\includegraphics[width=0.68\columnwidth]{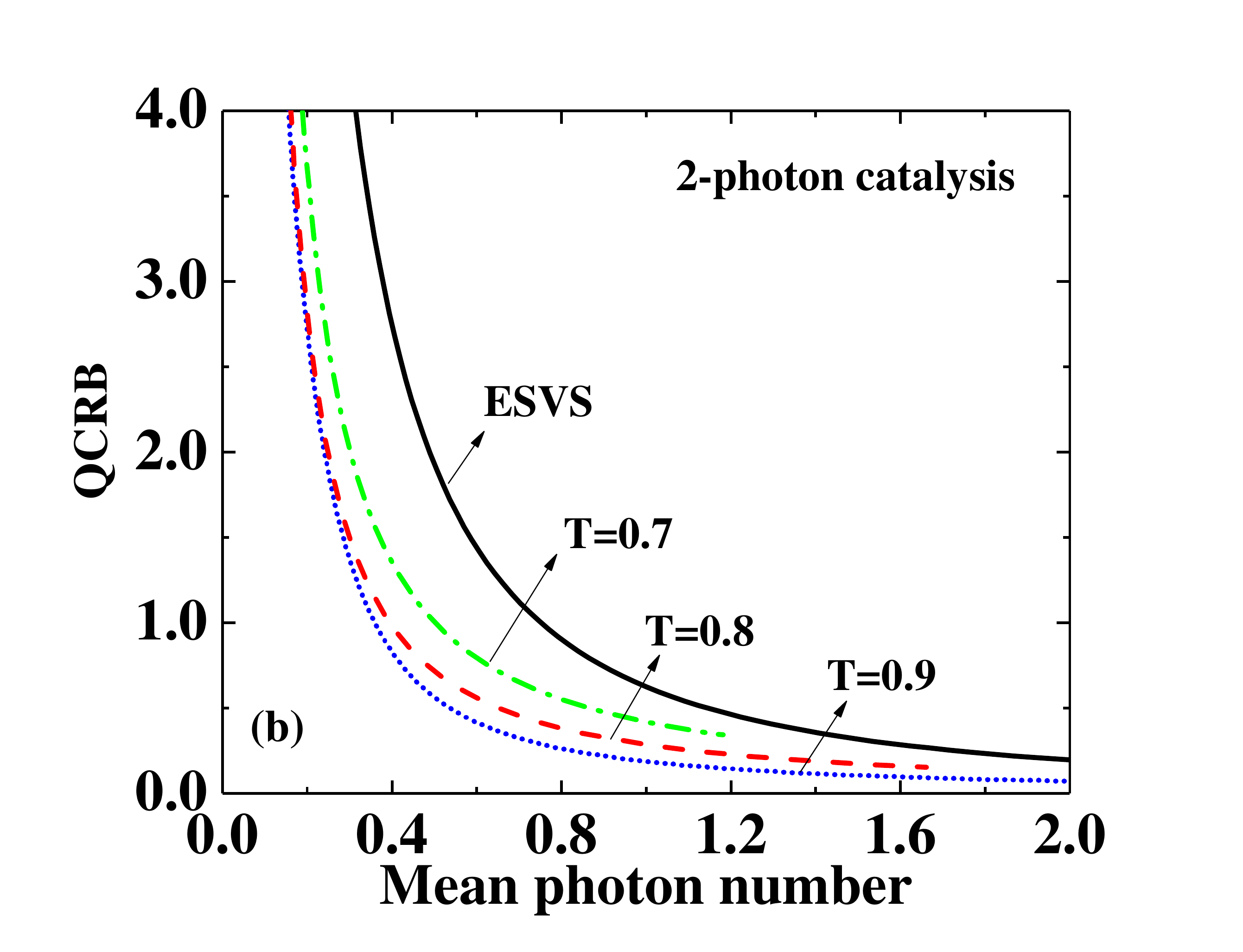}%
\includegraphics[width=0.68\columnwidth]{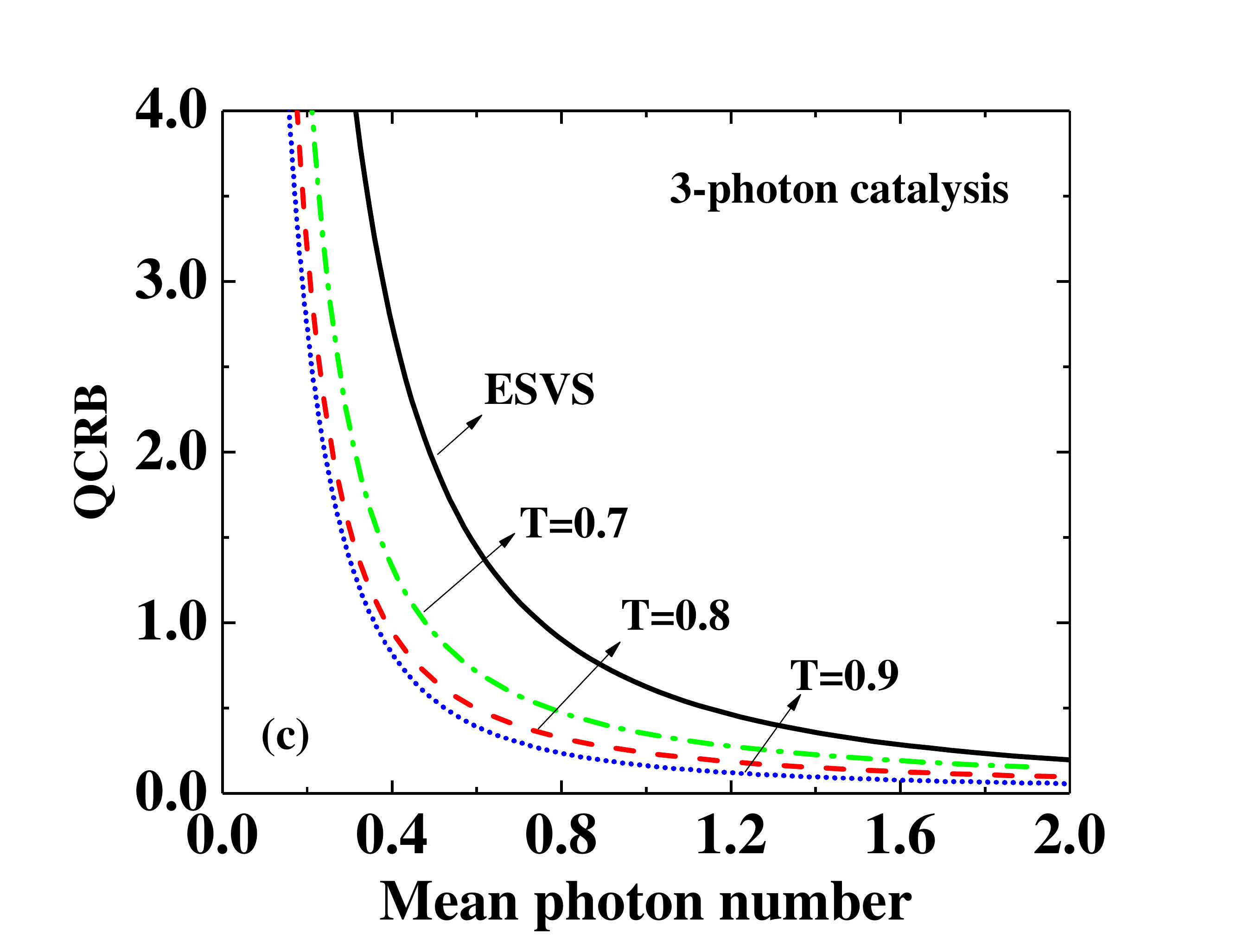}
\caption{{}(Color online) The QCRB as a function of mean photon number for
the ESVS (black) and the MPCESs with $T=0.9$ (blue), $T=0.8$ (red), and $%
T=0.7$ (green). (a) the case of single-photon catalytic, (b) the case of
two-photon catalytic, (c) the case of three-photon catalytic.}
\end{figure*}
\begin{figure*}[tbp]
\label{Fig4} \centering \includegraphics[width=0.6\columnwidth]{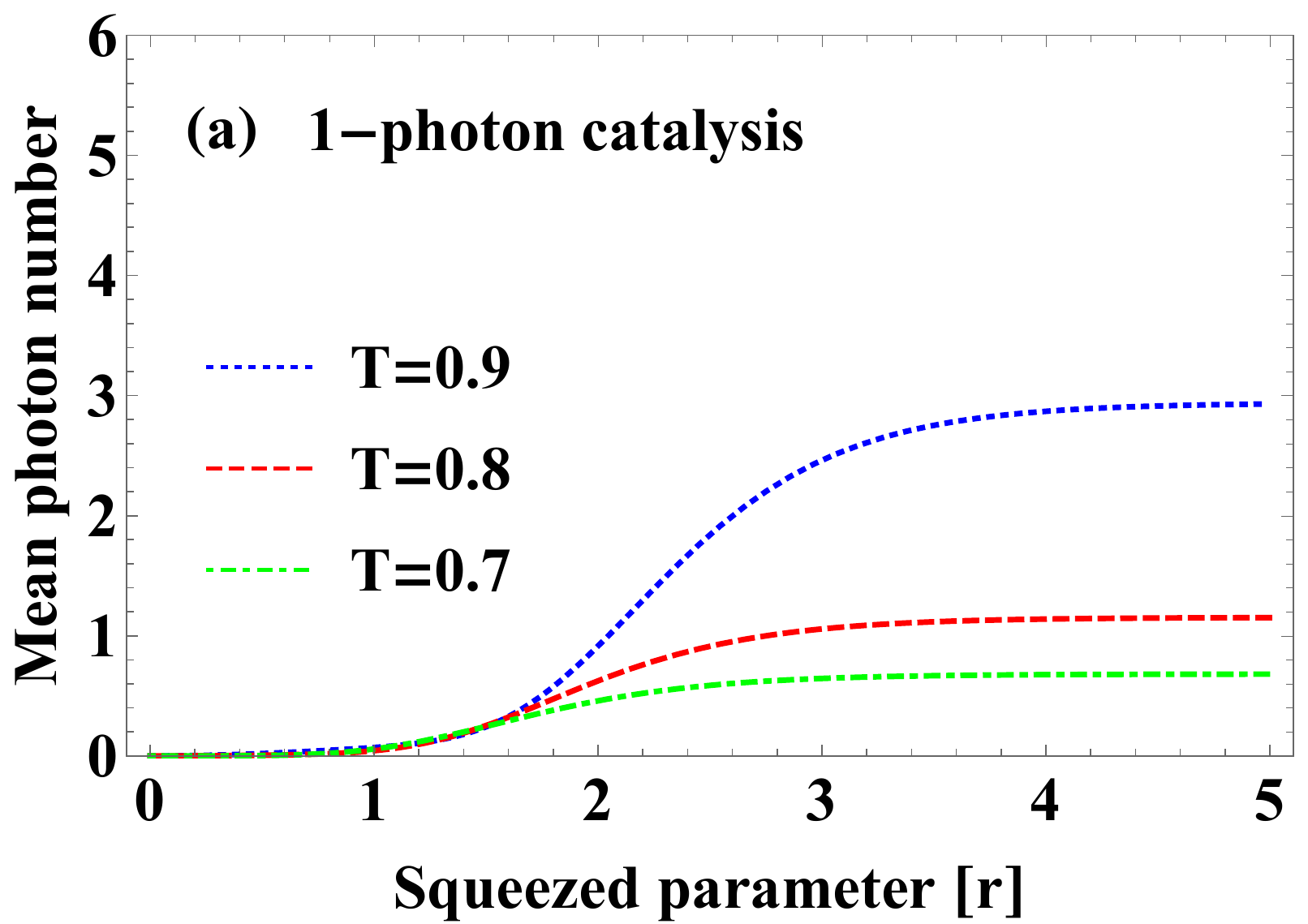} %
\includegraphics[width=0.6\columnwidth]{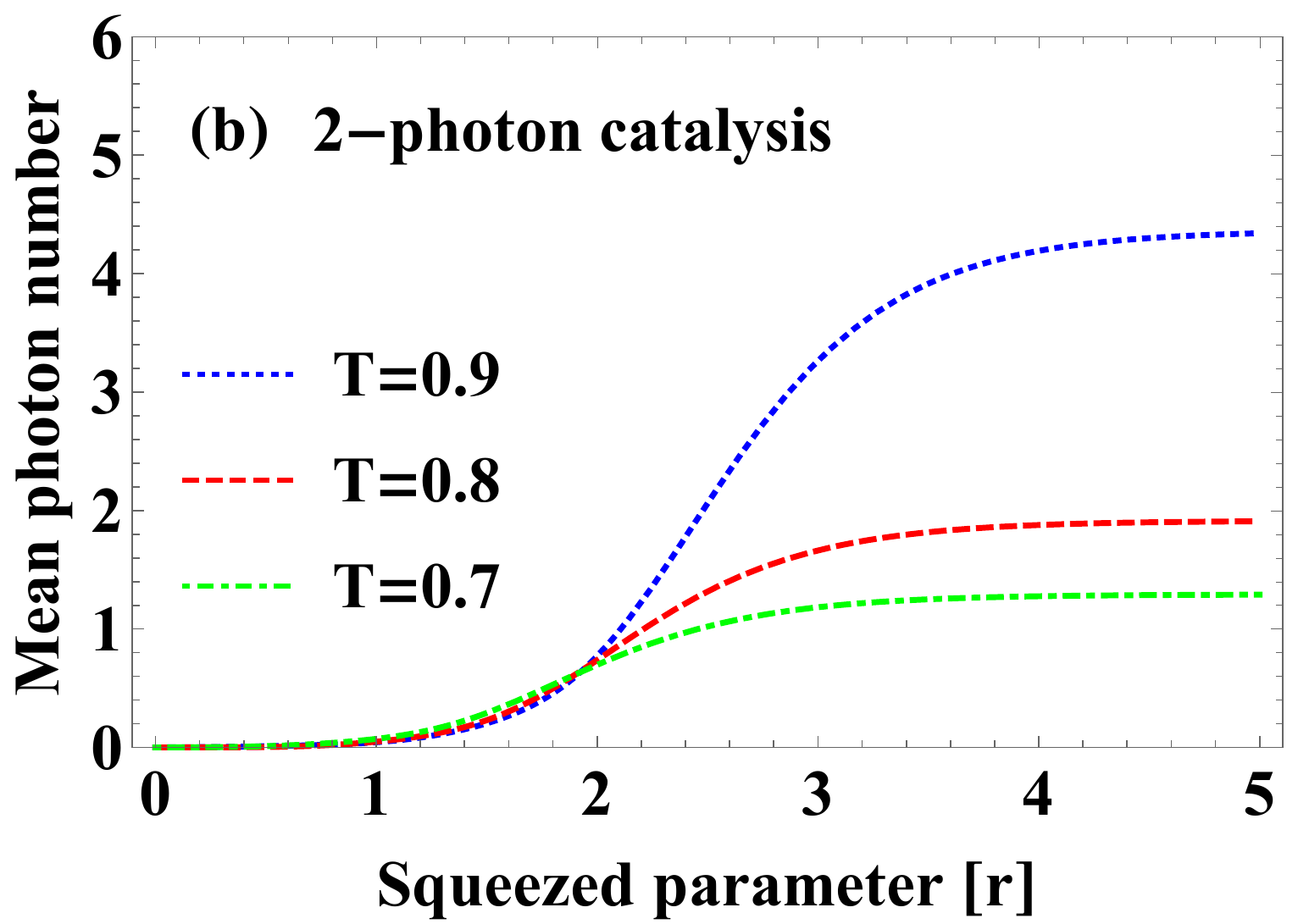}\includegraphics[width=0.6%
\columnwidth]{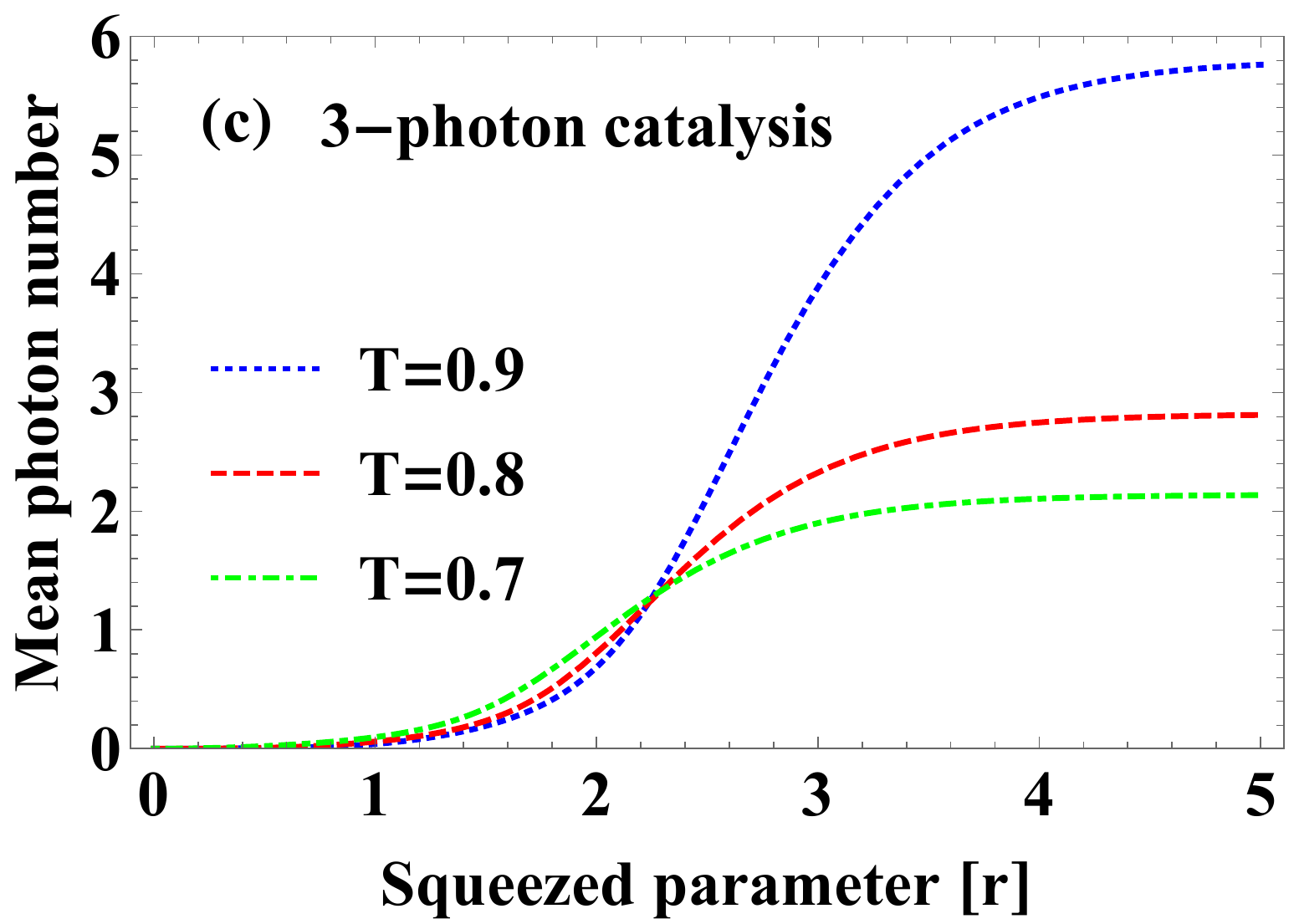}
\caption{{}(Color online) The mean photon number as a function of mean
photon number for the ESVS (black) and the MPCESs with $T=0.9$ (blue), $%
T=0.8 $ (red), and $T=0.7$ (green). (a) the case of single-photon catalytic,
(b) the case of two-photon catalytic, (c) the case of three-photon
catalytic. }
\end{figure*}

To elaborate the advantages of the MECSVS as inputs of the multi-arm
interferometer, we plot the $\left\vert \Delta \widehat{\mathbf{%
\theta }}\right\vert _{QCRB}^{2}$ as a function of the mean photon number
for several different catalytic photon numbers $n\in \{1,2,3\}$, as
shown in Fig. 3  where $d=5$ and $%
T=0.7,0.8,0.9$. As a comparison, the black solid line
corresponds to the result using the multi-mode ESVS case. It is clearly seen that the QCRBs in the case using MECSVS as inputs are obviously lower than those using the normal ESVS for all three catalytic photon  numbers (i.e., $n=1,2$ or $3$) especially when the average photon number is small. This indicates that by catalyzing the SSVS before inputting into the interferometer we can significantly improve the phase detection precision. The QCRB is the lowest when $T=0.9$ comparing with $T=0.7$ and $T=0.8$ for all three catalytic photon numbers.  We also note that the mean photon number of the MECSVS increases with the squeezing parameter $r$ but is saturated when $r$ is large enough (see Fig. 4). The maximum reachable mean photon number increases when $T$ or the catalysis photon number is larger.


\section{The QCRB of the multiparameter estimation with photon losses}

\begin{figure}[tbp]
\label{Fig5} \centering \includegraphics[width=7cm]{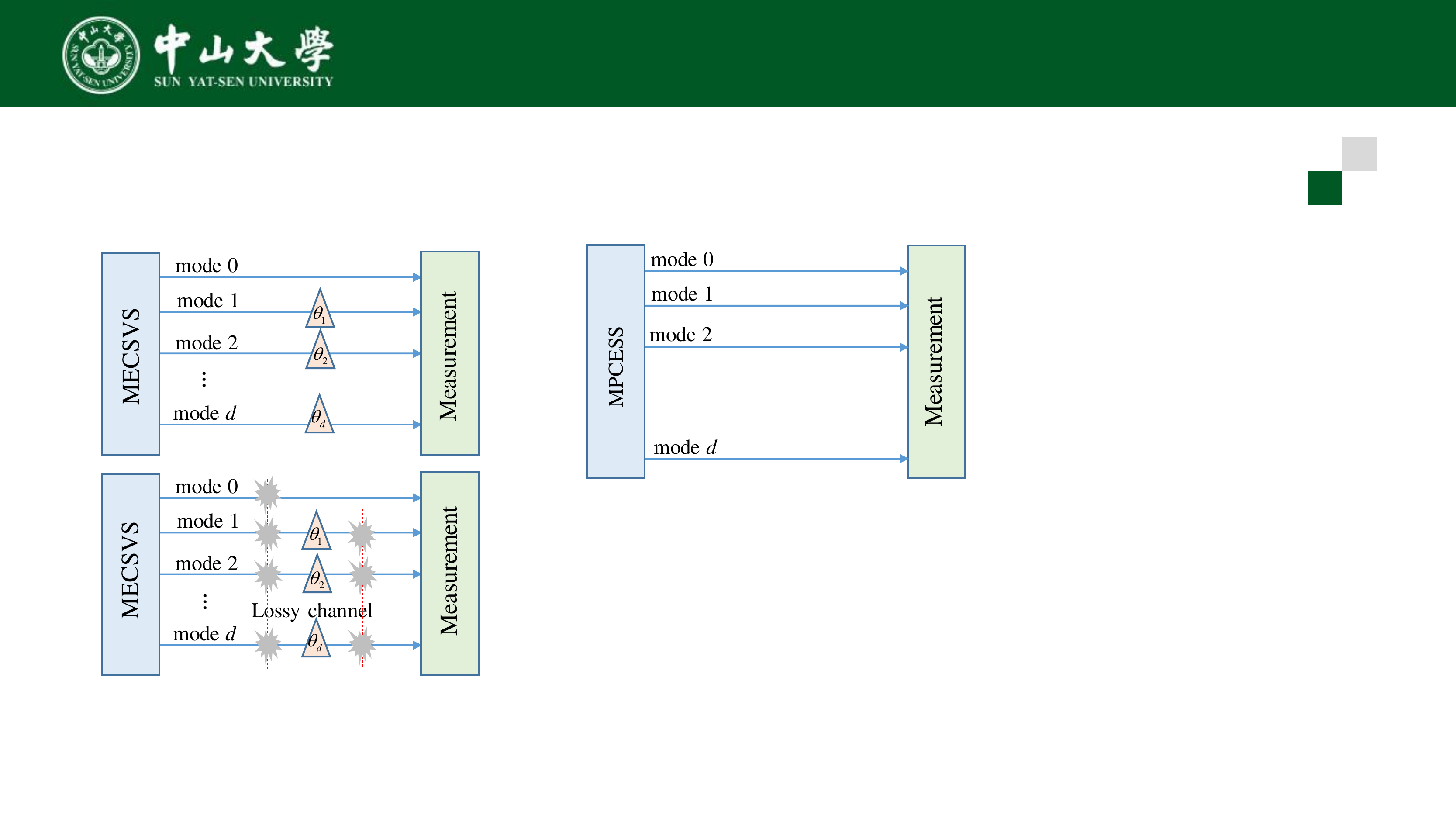}
\caption{{}(Color online) Schematic of multiparameter estimation with d
phase shifts under the photon losses.}
\end{figure}
\begin{figure*}[tbp]
\label{Fig6} \centering \includegraphics[width=0.9\columnwidth]{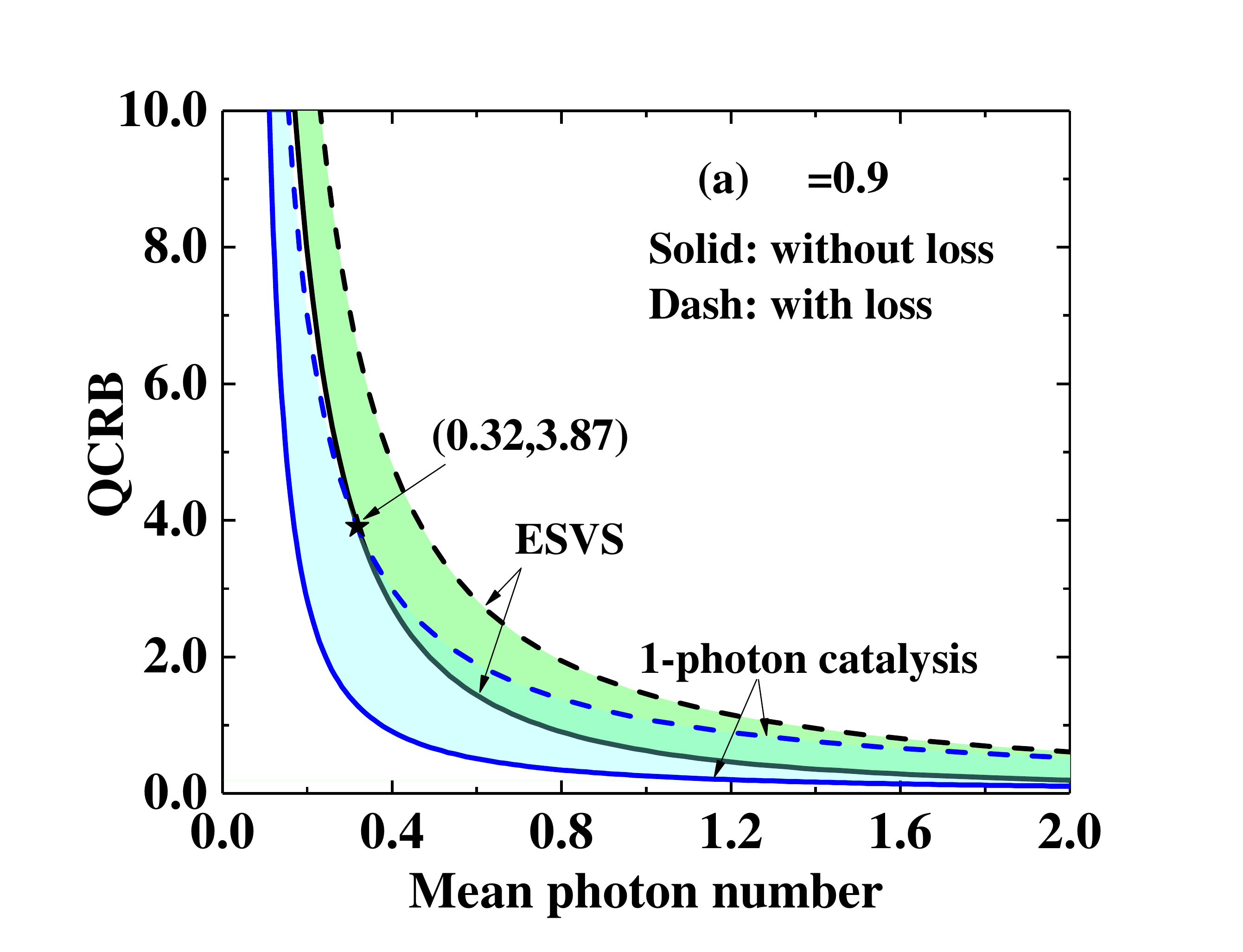} %
\includegraphics[width=0.9\columnwidth]{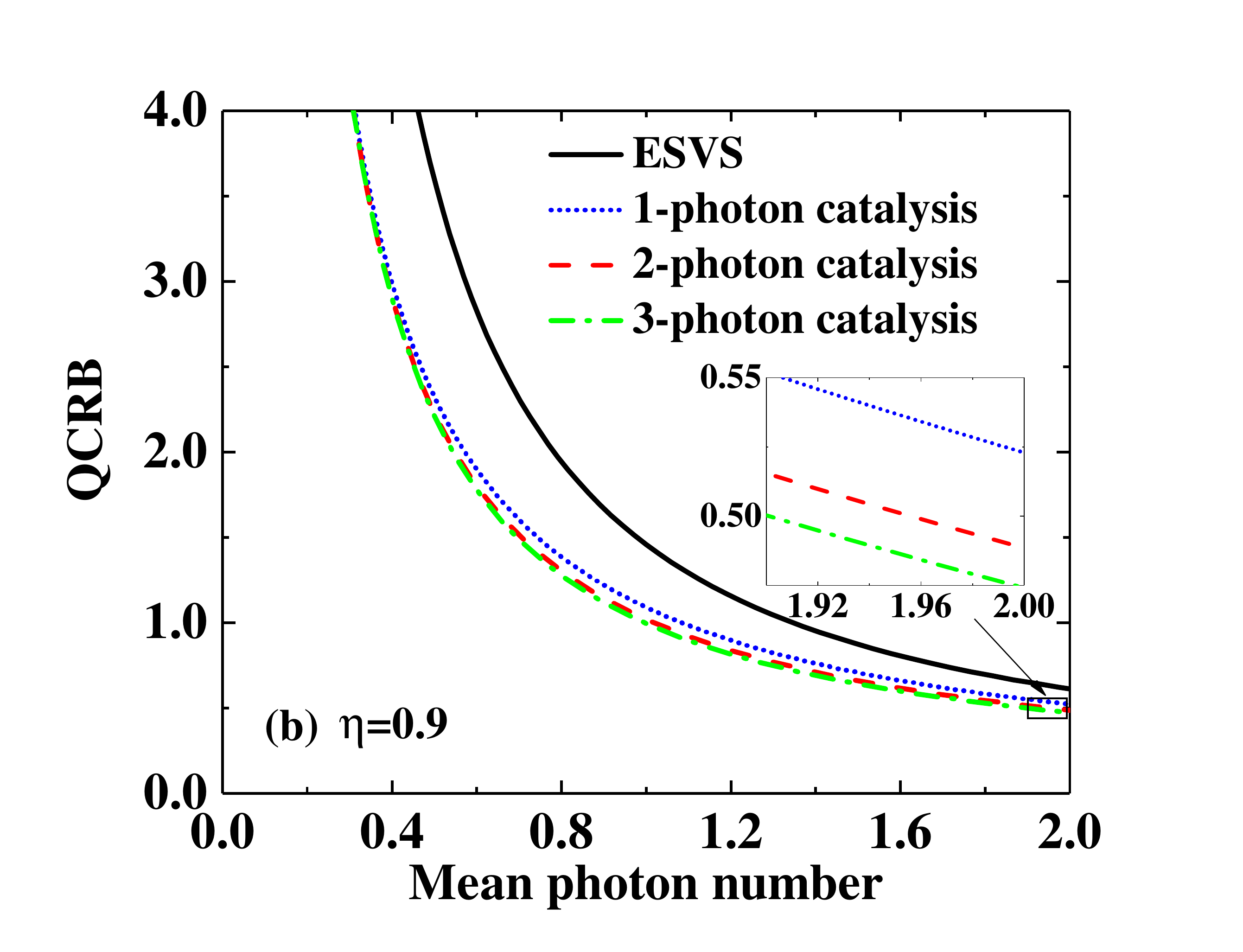}
\caption{{}(Color online) In the presence of photon losses, the phase
uncertainty $\left \vert \Delta \protect\widehat{\mathbf{\protect\theta }}%
\right
\vert _{QCRB}^{2}$ as a function of the mean photon number. (a)
Comparison of the ideal case and the loss case for the ESVS and the
single-photon catalysis operation. (b) Comparison of the phase uncertainty $%
\left \vert \Delta \protect\widehat{\mathbf{\protect\theta }}\right \vert
_{QCRB}^{2}$ with the number of the photon catalysis $n=1,2,$ and $3$, and
the ESVS (black dashed line) for 10\% photon loss ($\protect\eta $ $=0.9$).}
\end{figure*}
\begin{figure*}[tbp]
\label{Fig7} \centering \includegraphics[width=0.68\columnwidth]{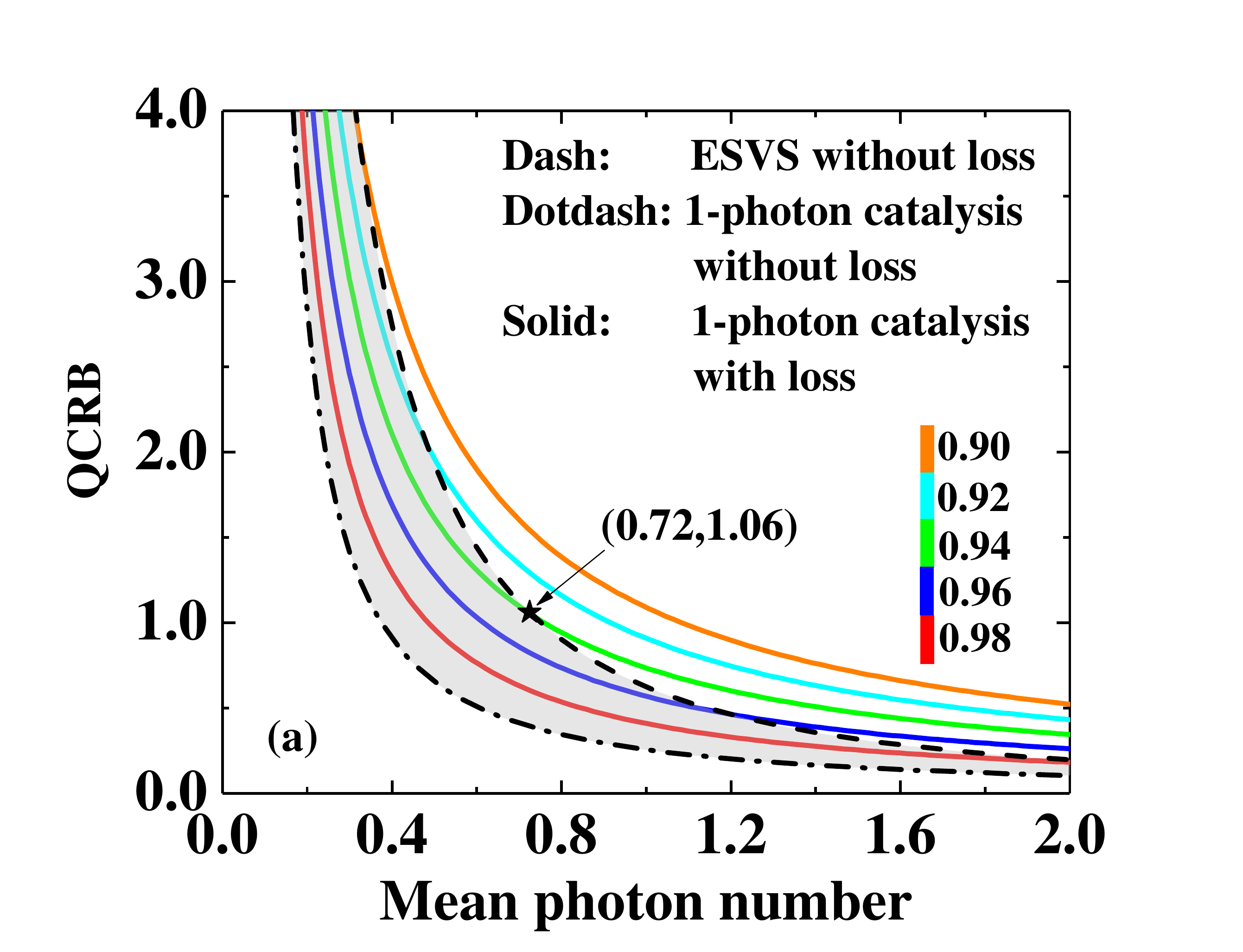} %
\includegraphics[width=0.68\columnwidth]{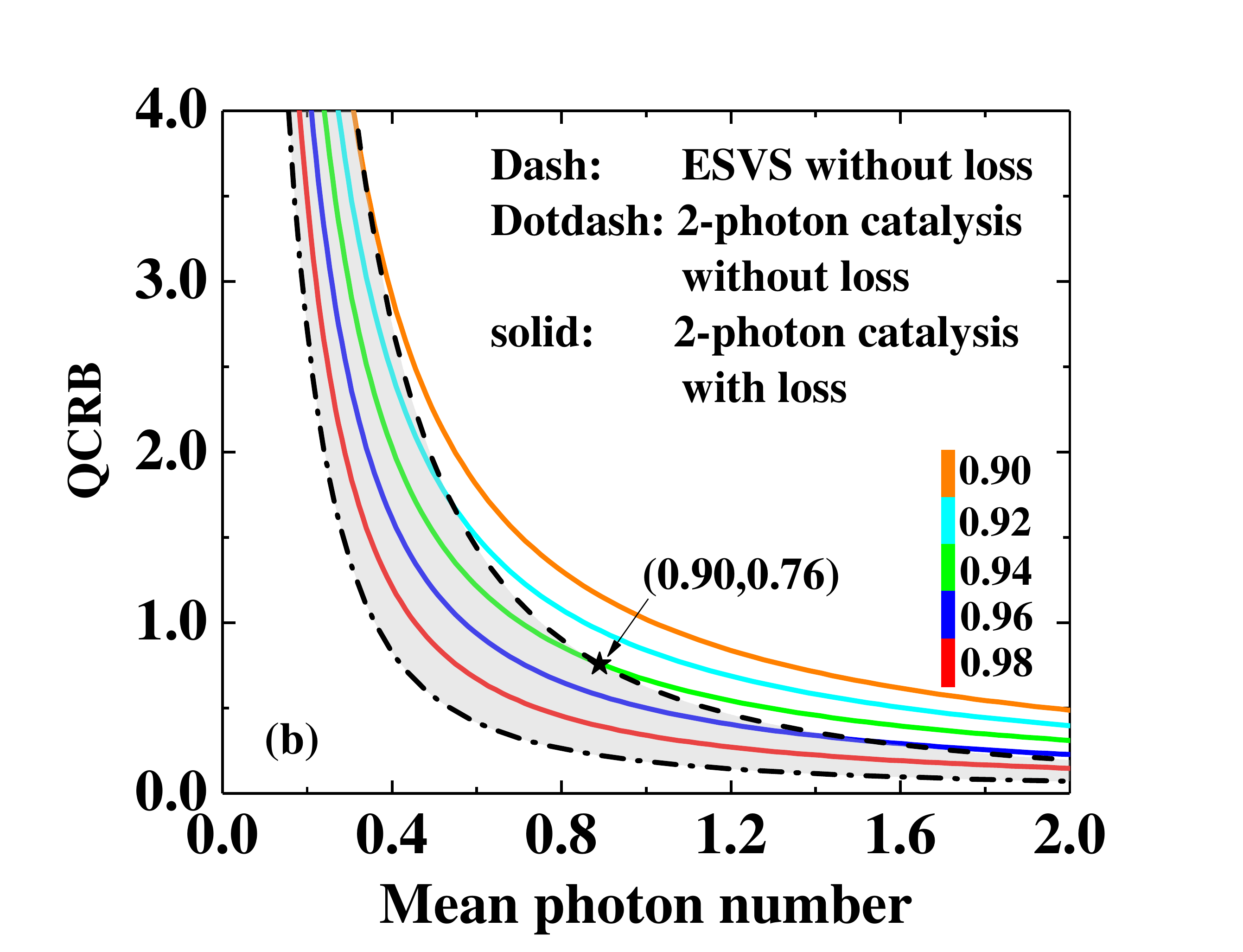}%
\includegraphics[width=0.68\columnwidth]{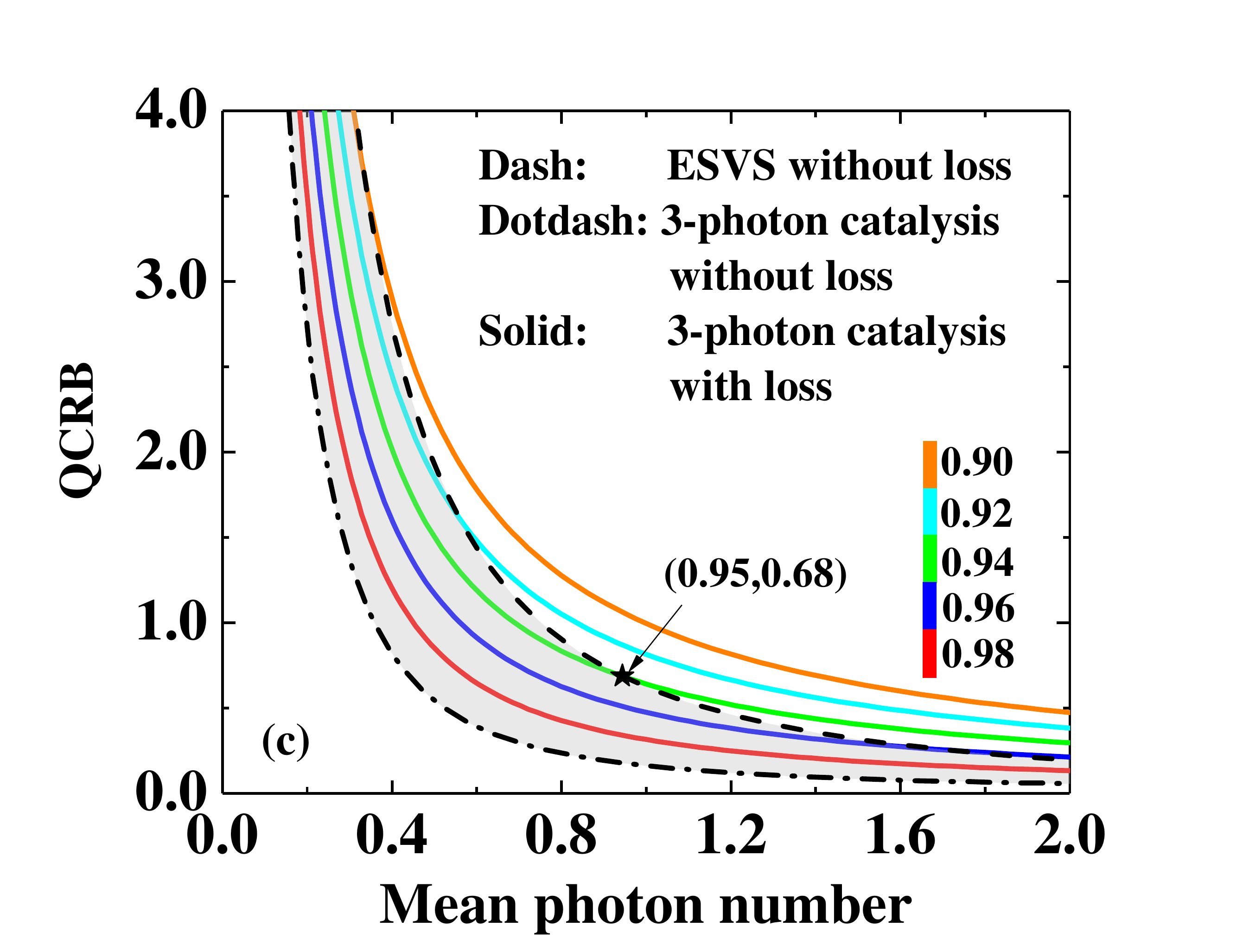}
\caption{{}(Color online) The QCRB as the function of the mean photon number
in the presence of photon losses. The Dot line represents the ideal ESVS,
the Dotdash line represents the single-photon catalytic ESVS. the colored
line represents the QCRB corresponding to different dissipation factors.}
\end{figure*}

Under the realistic environment, photon losses and phase diffusions are usually unadvoidable, which can affect the ultimate precision limit of the optical interferometry. Here we focus on studying the QCRB of multiparameter estimation with photon
losses, as shown in Fig. 5. To describe the influences of noise environment
on the phase probing system $S$, additional degrees of freedom should be
introduced, i.e., the environment denoted as $E$. Here we assume that the environment is in the vacuum state which is a reasonable assumption for the optical regime. In such circumstance, the initial state of the system and the environment are separable and can be written as $\left\vert \Psi
\right\rangle _{S}\left\vert \mathbf{0}\right\rangle _{E}$. The evolution of the whole combined system is unitary and can be denoted as $\widehat{U}%
^{S+E}\left( \mathbf{\theta }\right) $. Thus, the final state before the measurement can be expressed as \cite{61}
\begin{equation}
\left\vert \Psi \left( \mathbf{\theta }\right) \right\rangle _{S+E}\text{=}%
\widehat{U}^{S+E}\left( \mathbf{\theta }\right) \left\vert \Psi
\right\rangle _{S}\left\vert \mathbf{0}\right\rangle _{E}\text{=}\sum_{%
\mathbf{k}}\widehat{\Pi }_{\mathbf{k}}\left( \mathbf{\theta }\right)
\left\vert \Psi \right\rangle _{S}\left\vert \mathbf{k}\right\rangle _{E},
\label{19}
\end{equation}
where $\widehat{U}^{S+E}\left( \mathbf{\theta }\right) $=$%
U_{0}^{S_{0}+E_{0}}U_{1}^{S_{1}+E_{1}}\left( \theta _{1}\right) $....$%
U_{d}^{S_{d}+E_{d}}\left( \theta _{d}\right) $ and the environment vacuum state $\left\vert \mathbf{0}%
\right\rangle _{E}$=$\left\vert \mathbf{0}\right\rangle _{E_{0}}\left\vert
\mathbf{0}\right\rangle _{E_{1}}$....$\left\vert \mathbf{0}\right\rangle
_{E_{d}}$. In the second identity, $\left\vert \mathbf{k}\right\rangle _{E}$=$\left\vert
{k_{0}}\right\rangle _{E_{0}}\left\vert {k_{1}}\right\rangle _{E_{1}}$...$\left\vert
{k_{d}}\right\rangle _{E_{d}}$ are orthonormal states of the environment, and $%
\widehat{\Pi }_{\mathbf{k}}\left( \mathbf{\theta }\right) $=$\widehat{\Pi }%
_{k_{0}}\left( \theta _{0}\right) \otimes \widehat{\Pi }_{k_{1}}\left(
\theta _{1}\right) $....$\otimes \widehat{\Pi }_{k_{d}}\left( \theta
_{d}\right) $ is the direct product of all kraus operator $\widehat{\Pi }%
_{k_{j}}\left( \theta _{j}\right) $ \cite{62,63}, each of which is defined as%
\begin{equation}
\widehat{\Pi }_{k_{l}}\left( \theta _{l}\right) \text{=}_{E_{l}}\left\langle
k\right\vert \widehat{U}_{l}^{S_{l}+E_{l}}\left( \theta _{l}\right)
\left\vert 0\right\rangle _{E_{l}}.  \label{20}
\end{equation}%

From Eq. (\ref{19}), we can then calculate the QFI matrix of the whole system including the noisy environment and the QFI matrix elements is given by
\begin{equation}
\digamma (\mathbf{\theta ,}\widehat{\Pi }_{\mathbf{k}}\left( \mathbf{\theta }%
\right) )_{lj}=4(\left\langle \widehat{\pounds }^{\left( lj\right)
}\right\rangle -\left\langle \widehat{\Gamma }^{\left( l\right)
}\right\rangle \left\langle \widehat{\Gamma }^{\left( j\right)
}\right\rangle ),  \label{21}
\end{equation}%
where $\left\langle \cdot \right\rangle $ standing for $\left\langle \Psi
\right\vert \cdot \left\vert \Psi \right\rangle $ and%
\begin{equation}
\widehat{\Gamma }^{\left( l\right) }=\sum_{k_{l}}i\frac{d\widehat{\Pi }%
_{k_{l}}^{\dagger }\left( \theta _{l}\right) }{d\theta _{l}}\widehat{\Pi }%
_{k_{l}}\left( \theta _{l}\right) ,  \label{22}
\end{equation}%
\begin{equation}
\widehat{\pounds }^{\left( lj\right) }=\left\{
\begin{array}{c}
\sum_{k_{l}}\frac{d\widehat{\Pi }_{k_{l}}^{\dagger }\left( \theta
_{l}\right) }{d\theta _{l}}\frac{d\widehat{\Pi }_{k_{l}}\left( \theta
_{l}\right) }{d\theta _{l}},l=j \\
\\
\widehat{\Gamma }^{\left( l\right) }\widehat{\Gamma }^{\left( j\right)
},l\neq j%
\end{array}%
\right. .  \label{23}
\end{equation}%

As an example of the noise environment, the photon loss can be
simulated using an optical beam splitter with a transmissivity $\eta _{l}$ ($%
\eta _{l}=1$ corresponds to lossless case, and $\eta _{l}=0$ corresponds to complete photon
loss). From Eqs. (\ref{20}), (\ref{22}) and (\ref{23}), we can obtain $\widehat{\Gamma }^{\left(
l\right) }=\mu _{l}\widehat{N}_{l}$ and $\widehat{\pounds }^{\left( ll\right)
}=\mu _{l}\widehat{N}_{l}^{2}+v_{l}\widehat{N}_{l}$ when $l=j$ (and $\widehat{%
\pounds }^{\left( lj\right) }=\widehat{\Gamma }^{\left( l\right) }\widehat{%
\Gamma }^{\left( j\right) }$ when $l\neq j$), with $\mu _{l}=1-\left(
1+\varepsilon _{l}\right) \left( 1-\eta _{l}\right) $ and $v_{l}=\left(
1+\varepsilon _{l}\right) ^{2}\eta _{l}\left( 1-\eta _{l}\right) $. As shown
in Fig. 5, if such photon losses exist in the each mode of the multi-arm
interferometer, a series of Kraus operators in each mode is given by \cite%
{36}%
\begin{equation}
\widehat{\Pi }_{k_{l}}\left( \theta _{l}\right) =\sqrt{\frac{\left( 1-\eta
_{l}\right) ^{k_{l}}}{k_{l}}}e^{i\theta _{l}\left( \widehat{N}%
_{l}-\varepsilon _{l}k_{l}\right) }\eta _{l}^{\frac{\widehat{N}_{l}}{2}%
}a_{l}^{k_{l}},  \label{24}
\end{equation}%
where $\varepsilon _{l}$ is an arbitrary real number.  $\varepsilon _{l}$ need to be
optimized to make the lower bound $\digamma (\mathbf{\theta ,}\widehat{\Pi }%
_{\mathbf{k}})$ as tight as possible. According to Eqs. (\ref{21}) and (%
\ref{24}), the lower bound for the optimal precision of multiparameter
estimation is given by \cite{64,65}
\begin{equation}
\text{cov}\left( \theta \right) \geq \frac{1}{\digamma (\mathbf{\theta ,}\widehat{%
\Pi }_{\mathbf{k}})}.  \label{25}
\end{equation}

Based on Eq. (\ref{25}), let us take the MECSVS as the input of multi-arm
interferometer. In this case, all (off-) diagonal elements of $\digamma (%
\mathbf{\theta ,}\widehat{\Pi }_{\mathbf{k}})$ are the same, denoted as  $\digamma _{d}$ ($%
\digamma _{o}$). For simplicity, here
we can make a reasonable assumption that $\eta _{l}=\eta $ and $\varepsilon
_{l}=\varepsilon $ since all modes are symmetric for the probe state and then we can obtain
\begin{eqnarray}
\digamma _{d} &=&4(\mu _{l}^{2}\left\langle \Delta \widehat{N}%
_{l}^{2}\right\rangle +v_{l}\left\langle \widehat{N}_{l}\right\rangle ),
\label{26} \\
\digamma _{o} &=&4[\mu _{l}^{2}(\left\langle \widehat{N}_{l}\widehat{N}%
_{j}\right\rangle -\left\langle \widehat{N}_{l}\right\rangle \left\langle
\widehat{N}_{j}\right\rangle )],  \label{27}
\end{eqnarray}%
where $\left\langle \Delta \widehat{N}_{l}^{2}\right\rangle =\left\langle
\widehat{N}_{l}^{2}\right\rangle -\left\langle \widehat{N}_{l}\right\rangle
^{2}$ and $\left\langle \widehat{N}_{l}\widehat{N}_{j}\right\rangle =0,$ $%
\left\langle \widehat{N}_{l}\right\rangle \left\langle \widehat{N}%
_{j}\right\rangle =\left\langle \widehat{N}_{l}\right\rangle ^{2}$ for the
MECSVS. Then we can calculate the analytical expression of the QCRB with MECSVS as inputs of the multi-arm interferometer
under the photon losses \cite{36}%
\begin{equation}
\left\vert \Delta \widehat{\mathbf{\theta }}\right\vert _{QCRB}^{2}=\max_{%
\widehat{\Pi }_{\mathbf{k}}}\text{Tr}[\digamma ^{-1}(\mathbf{\theta ,}%
\widehat{\Pi }_{\mathbf{k}})],  \label{28}
\end{equation}%
where
\begin{equation}
\text{Tr}[\digamma ^{-1}(\mathbf{\theta ,}\widehat{\Pi }_{\mathbf{k}})]\text{%
=}\frac{d-1}{\digamma _{d}-\digamma _{o}}+\frac{1}{\digamma _{d}+\left(
d-1\right) \digamma _{o}}.  \label{29}
\end{equation}

In Fig. 6(a), we compare the QCRBs of the ideal (solid lines) and the photon-loss (dashed lines) cases with different input states (i.e., the ESVS and the single-photon catalyzed state) where the strength of the photon loss is chosen to be $\eta =0.9$. We can see that with photon loss, the QCRBs increase for both inputs. However, we can see that the QCRB of using the single-photon catalyzed state is still lower than that using the ESVS case which indicates that the single-photon catalyzed state still has better performance than the ESVS case under the photon losses. We also find that when the mean photon number is less than $0.32$, the QCRB for the MECSVS with $n=1$ with photon losses can be still lower than that for the multi-mode ESVS without photon losses.  In addition, we also consider the QCRB with the input MECSVS for several different catalytic photon numbers $n\in\{1,2,3\},$ as shown in Fig. 6(b) where  $\eta =0.9$. It is found that, the QCRBs for all the three catalysis photon numbers are about the same with photon losses and all of them perform better than that with the ESVS input. 

We also investigate the QCRB for different strength of the photon loss which are shown in Fig. 7 where we set $T=0.9$. From the three subfigures, we can see that with the decrease of $\eta \in \{1,0.98,0.96,0.94,0.92,0.90\},$ the QCRB\ for the MECSVS increases, meaning that the precision of multiparameter estimation decreases with the increase of photon-loss intensity. More interestingly, wih the increase of $n=1,2,3$, the usage of the MECSVS with respect to the QCRB canbe superior to that of the multi-mode ESVS at a certain range of the mean
photon number. To be more specific, for $\eta =0.94$ (green line), when the
mean photon number is less than $0.72,0.90$ and $0.95$, the QCRB for the
MECSVS of $n=1,2,3$ with photon losses can be lower than that for the multi-mode
ESVS without photon losses, and such a phenomenon is also true when $\eta =0.96$
and $0.98$. Thus, the results above clearly show that using the MECSVS can improve the phase estimation precision over the ESVS in both the noisy and noisy-free cases.



\begin{figure*}[htp]
\label{Fig8} \centering \includegraphics[width=0.9\columnwidth]{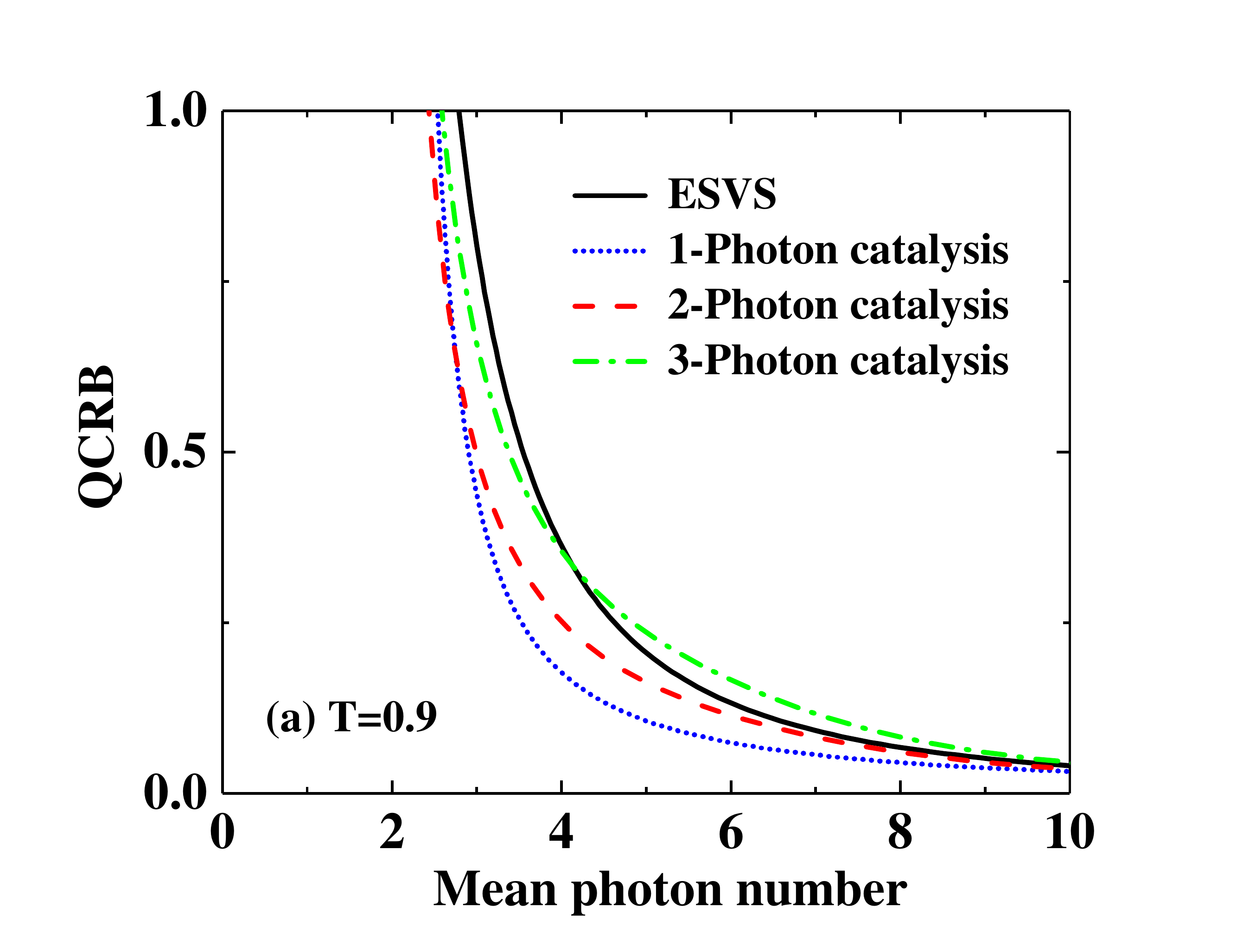} %
\includegraphics[width=0.9\columnwidth]{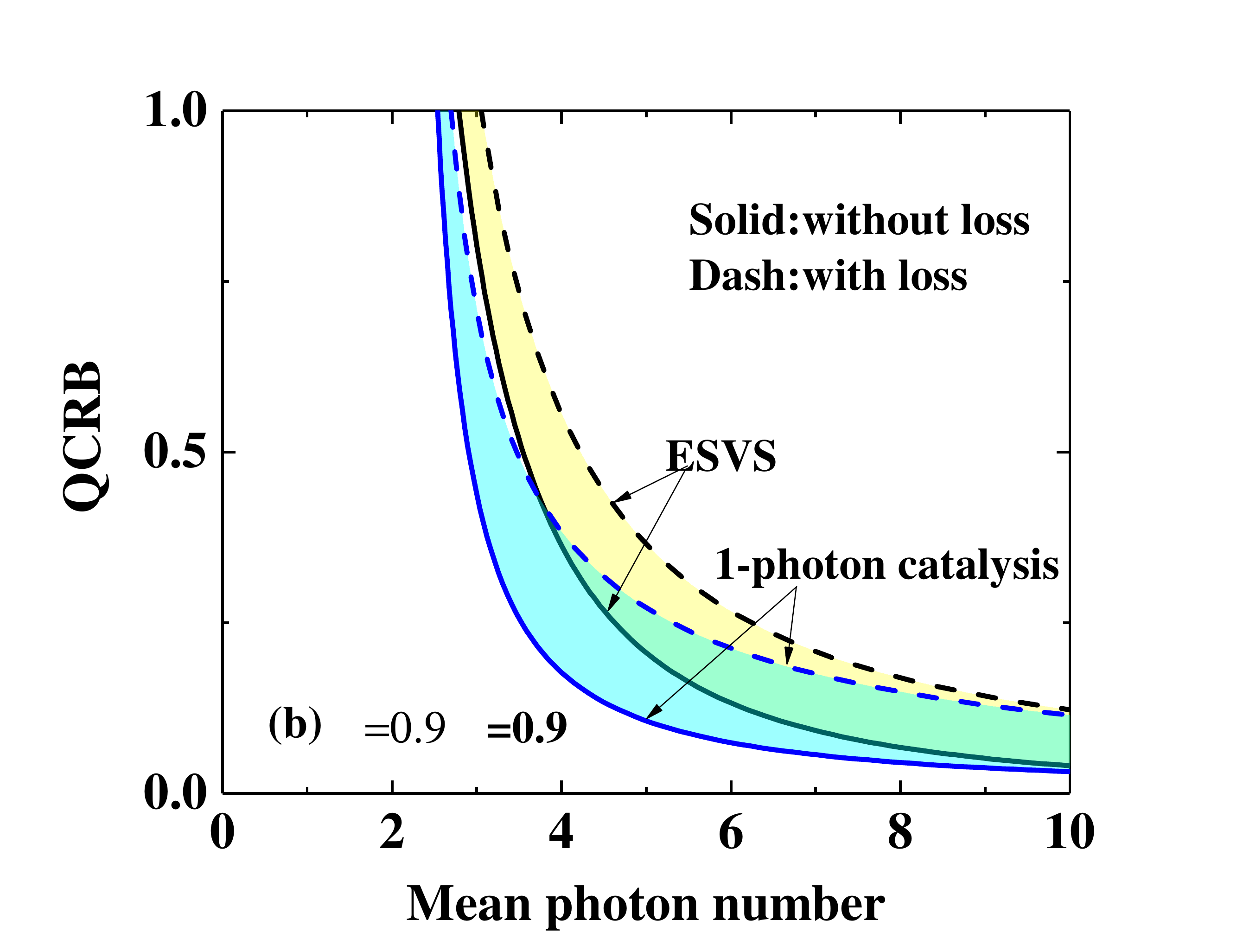}
\caption{{}(Color online) The QCRB as a function of the mean photon number (a) for given $T=0.9$ and different catalysis photon number $n=1,2,3$ in ideal case and (b) for given $T=0.9$ and $\eta =0.9$ with photon losses. }
\end{figure*}

\section{The QCRB of the other multi-mode entangled states}

In Sec. II, we show that if the measurement results for $D_1$ and $D_2$ detectors are zero and one photon, respectively, antisymmetric instead of symmetric superposition of vacuum and  ESVS is obtained. In this section, we show that this antisymmetric state can be also used for improving the phase estimation precision. As an example, we assume that all the $D_1$ and $D_2$ detectors detect zero and one photon, respectively. In this case, the output state for the 6 modes example can be calculated as%
\begin{eqnarray}
\left\vert \Psi ^{\prime }\right\rangle  &=&\widetilde{%
\mathbb{N}
}^{\prime }(\left\vert \xi \right\rangle _{0}\left\vert 0\right\rangle
_{1}\left\vert 0\right\rangle _{2}\left\vert 0\right\rangle _{3}\left\vert
0\right\rangle _{4}\left\vert 0\right\rangle _{5} \notag \\ &&-\left\vert 0\right\rangle
_{0}\left\vert \xi \right\rangle _{1}\left\vert 0\right\rangle
_{2}\left\vert 0\right\rangle _{3}\left\vert 0\right\rangle _{4}\left\vert
0\right\rangle _{5} \notag \\ &&-\left\vert 0\right\rangle _{0}\left\vert 0\right\rangle
_{1}\left\vert \xi \right\rangle _{2}\left\vert 0\right\rangle
_{3}\left\vert 0\right\rangle _{4}\left\vert 0\right\rangle _{5} \notag \\ &&+\left\vert
0\right\rangle _{0}\left\vert 0\right\rangle _{1}\left\vert 0\right\rangle
_{2}\left\vert \xi \right\rangle _{3}\left\vert 0\right\rangle
_{4}\left\vert 0\right\rangle _{5}  \notag \\
&&-\left\vert 0\right\rangle _{0}\left\vert 0\right\rangle _{1}\left\vert
0\right\rangle _{2}\left\vert 0\right\rangle _{3}\left\vert \xi
\right\rangle _{4}\left\vert 0\right\rangle _{5} \notag \\ &&+\left\vert 0\right\rangle
_{0}\left\vert 0\right\rangle _{1}\left\vert 0\right\rangle _{2}\left\vert
0\right\rangle _{3}\left\vert 0\right\rangle _{4}\left\vert \xi
\right\rangle _{5}), \label{32} 
\end{eqnarray}
where $\widetilde{%
\mathbb{N}
}^{\prime }=1/\sqrt{6\left\langle \xi ^{\prime }\right\vert \left. \xi
^{\prime }\right\rangle -6\left\vert \left\langle 0\right\vert \left. \xi
^{\prime }\right\rangle \right\vert ^{2}}$ is the normalization coefficient.
Based on Eq. (\ref{32}), we can calculate the QCRB of multiparameter estimations with and without photon losses.

In order to more see whether this entangled state [Eq. (\ref{32})] can be  also used to improve the precision of multiphase estimation or not, we plot the QCRB as a function of the mean photon number for given $T=0.9$ and different catalysis photon number $n=1,2,3$ in the ideal case [see Fig. 8(a)] and in the case with photon losses with $\eta =0.9$ [see Fig. 8(b)]. For comparison, the QCRBs of ESVS as input state with and without photon losses are also shown in the figures. From Fig. 8(a), we can see that, for the entangled state given in Eq. (\ref{32}), single-photon catalysis and two-photon catalysis have obvious advantages in improving the precision of multiphase estimation. However, when the number of catalytic photons $n\geq 3$, the QCRB of the catalytic entangled states does not surpass that of the ESVS when the mean photon number is large (i.e, larger than 4.13 in this example). From Fig. 8(b), we can see that the QCRB in the case of single-photon catalysis is still lower than that in the ESVS case with photon losses. Hence, using the MECSVS in the antisymmetric case can also improve the phase estimation precision over using the ESVS.

\section{Conclusion}

In this paper, we proposed a method to prepare the MECSVS and then shown that using MECSVS as input of a multi-arm optical interferometer, we can improve the QCRB of the multi-phase measurement over that using the corresponding usual ESVS in both cases with and without photon losses.  We also compared the performance of the single-, two-, and three-photon catalysis states. The results shown that in both the cases with and without photon losses for the symmetric MECSVS, the three-photon catalysis entangled state have better performance than the other two, which indicates that multi-photon catalysis has more resilient to the environment noises in this case. Additionally, for the antisymmetric MECSVS, it turns out that the single-photon catalyzed entangled state gives the best QCRB, which implies that the MECSVS with antisymmetric case can be also used to improve the phase estimation precision. Our results can find important applications in the quantum metrology for multiparameter estimation.  

\section{Acknowledgements}
This work is supported by the Key-Area Research and Development Program of Guangdong Province (Grant No. 2018B030329001); the National Key R\&D Program of China (Grant No. 2021YFA1400800); Natural Science Foundations of Guangdong (Grant No. 2021A1515010039); National Natural Science Foundation of China (11964013); Major Discipline Academic and Technical Leaders Training Program of Jiangxi Province (20204BCJL22053).

\end{document}